\documentclass[twocolumn, switch]{article} 

\pdfoutput=1

\usepackage{preprint}

\usepackage[numbers,square]{natbib}
\usepackage[utf8]{inputenc}	
\usepackage[T1]{fontenc}	
\usepackage{xcolor}		
\usepackage[colorlinks = true,
            linkcolor = purple,
            urlcolor  = blue,
            citecolor = cyan,
            anchorcolor = black]{hyperref}	
\usepackage{booktabs} 		
\usepackage{nicefrac}		
\usepackage{microtype}		
\usepackage{lineno}		
\usepackage{float}			

\usepackage{lipsum}		

\usepackage{amsmath,amssymb,amsfonts}

\newcommand{\argmin}{\mathop{\rm arg~min}\limits}
\usepackage{amsthm}
\usepackage{amssymb}
\usepackage{amsmath,bm}
\usepackage{fixmath}
\usepackage{here}
\usepackage{algorithm}
\usepackage{algorithmicx}
\usepackage[noend]{algpseudocode}

\usepackage{fancyhdr}

\usepackage{newfloat}
\DeclareFloatingEnvironment[name={Supplementary Figure}]{suppfigure}
\usepackage{sidecap}
\sidecaptionvpos{figure}{c}

\usepackage[small,raggedright,compact]{titlesec}
\titlespacing\section{0pt}{1pt plus 1pt minus 1pt}{1pt plus 1pt minus 1pt}
\titlespacing\subsection{0pt}{1pt plus 1pt minus 1pt}{1pt plus 1pt minus 1pt}
\titlespacing\subsubsection{0pt}{1pt plus 1pt minus 1pt}{1pt plus 1pt minus 1pt}

\usepackage{tikz,xcolor,hyperref}

\definecolor{lime}{HTML}{A6CE39}
\DeclareRobustCommand{\orcidicon}{
	\begin{tikzpicture}
	\draw[lime, fill=lime] (0,0) 
	circle [radius=0.16] 
	node[white] {{\fontfamily{qag}\selectfont \tiny ID}};
	\draw[white, fill=white] (-0.0625,0.095) 
	circle [radius=0.007];
	\end{tikzpicture}
	\hspace{-2mm}
}
\foreach \x in {A, ..., Z}{\expandafter\xdef\csname orcid\x\endcsname{\noexpand\href{https://orcid.org/\csname orcidauthor\x\endcsname}
			{\noexpand\orcidicon}}
}

\title{{\bf Locally Self-Adjustive Smoothing for Measurement Noise Reduction \\ with Application to Automated Peak Detection}}

\usepackage{authblk}

\author[1*]{Keisuke Ozawa}
\author[2]{Tomoya Itakura}
\author[2]{Taisuke Ono}

\affil[1]{Research and Development, DENSO IT Laboratory, Shibuya, 150-0002, Tokyo, Japan}
\affil[*]{E-mail: ozawa.keisuke@core.d-itlab.co.jp}
\affil[2]{DENSO CORPORATION, 1-1 Showa-cho, Kariya, Aichi 448-8661, Japan}

\begin{document}

\twocolumn[ 
  \begin{@twocolumnfalse} 
  
\maketitle

\vspace{0.35cm}

  \end{@twocolumnfalse} 
] 


\section*{\centering {\bf Abstract}}
Smoothing is widely used approach for measurement noise reduction in spectral analysis. However, it suffers from signal distortion caused by peak suppression. A locally self-adjustive smoothing method is developed that retains sharp peaks and less distort signals. The proposed method uses only one parameter that determines global smoothness, while balancing the local smoothness using data itself. Simulation and real experiments in comparison with existing convolution-based smoothing methods indicate both qualitatively and quantitatively improved noise reduction performance in practical scenarios. We also discuss parameter selection and demonstrate an application for the automated smoothing and detection of a given number of peaks from noisy measurement data.

\section{Introduction}
Spectral data play a central role in chemical analysis.
Peak positions and their intensities proivde important information such as the bonding of molecules, crystallinity of solids, level of material degradation, and so on~\cite{shen2010effect, amin2015production, zhuo2015effect, perret2021high, llovet2022electron}. However, measurement data are affected by noise amplified in the derivatives of the data, which severely affect peak analysis. Noise reduction is often considered a fundamental data preprocessing step for improving the signal-to-noise ratio (SNR), accuracy of peak detection, and peak-intensity estimation. The noise handling has also gained considerable importance given the recent demands of data-driven development for conducting a high-throughput, automated analysis.

Smoothing is a popular noise reduction approach. Although smoothing introduces a bias in the processed data, chemically informative signals are smooth over data points, and fluctuations over a small interval can be attributed to noise. Frequency-based methods~\cite{horlick1972digital, kauppinen1982smoothing, shao2003wavelet} attribute high-frequency components as noise; however, thresholding the frequency depends on manual inspection. Further, peaks also have high-frequency components, and false peaks often remain or are amplified. Weighted-averaging-based methods apply convolution with various kernels~\cite{barton2022chemometrics, schmid2022and}. Some methods require calibration datasets for specific scenarios~\cite{chen2014recovery, bai2020denoising}. Among general purpose methods, Gaussian smoothing is a popular technique with a simple, band-limited convolution, and it is used in various fields. The Savitzyky–Golay method (SG)~\cite{savitzky1964smoothing} is the most popular smoothing method in chemical and biomedical analyses, wherein a data is locally approximated as a polynomial function, and each data point is replaced with the value of the function. SG is implemented as a convolution by fixing the degree of polynomials and the window size. However, SG often distorts data sensitively to the parameters. As an improved alternative, Eilers introduced the smoothing spline technique, originally developed by Whittaker~\cite{whittaker1922new}, for chemical analysis, and provided an efficient implementation for discrete problems. Eilers' perfect smoother (PS)~\cite{eilers2003perfect} formulates the smoothness regularized least squares problem using only one smoothness parameter. Existing studies have shown that PS is preferred over SG, which can invert the signal phase, and Gaussian smoothing, which can suppress peaks more strongly~\cite{schmid2022and}. Eilers suggested that penalizing second derivative achieves good results, which is closely related to enhancing the peaks from noisy data.

SG and PS are widely used for smoothing in chemical analysis and its related subjects. Though, SG and PS have a common source of signal distortion: Sharp peaks become suppressed when reducing noise, and some important peaks become less prominent and difficult to be detected. In terms of convolution, this is attributed to the weighted averaging with a fixed convolution that ignores the local smoothness of data. Wider window sizes can smooth data more strongly, while sharp peaks are averaged over their tails. To overcome this issue, Luo et al.~\cite{luo2021developing} proposed extracting the sharp peaks and then smoothing the data points within the remainder regions. Yao et al.~\cite{yao2021yield} addressed the peak suppression problem by selecting sharp peaks and linearly interpolating them with the subresolution measurement steps (in the spectral domain). Although these methods can maintain the sharp peaks while smoothing data, the peaks need to be empirically extracted before  smoothing.

In this study, we develop a locally self-adjustive smoothing that retains sharp peaks based on PS and demonstrate improved peak detection using the minima of second derivatives, without extracting the peaks before smoothing and any manual inspection. Local smoothness of data has also been considered in previous studies; however, they require tuning multiple parameters, slow iterated computations~\cite{urbas2011automated, barton2018algorithm}, and empirically determined regional segmentations~\cite{urbas2011automated, wang2013smoothing}. In contrast, the proposed method can be applied in a data-driven manner with a single filtering. The required parameter is one that controls the global smoothness of data as required for PS. From a practical standpoint, selecting the parameter is a crucial issue. PS provides an efficient, statistics-based~\cite{hastie1990generalized} parameter selection criterion based on the assumption that the noise is uncorrelated. However, it is empirically known that the selected parameter tends to be underestimated, and thus, manual inspection is still required in practice. To realize a fully automated application, we empirically modified the parameter selection criterion by considering the local smoothness of data, and we demonstrated that the underestimation issue is improved. Throughout simulation and real experiments, we consider different scenarios with various noise levels and sharp peaks, wherein show that the proposed method would be a better alternative to the existing methods.

\section{Methodology}
\subsection{Eilers' PS and the unbalanced local smoothness}
The problem is to estimate the underlying signal $\mathbf{x}^{\circ} \in \mathbb{R}^N$ from a noisy observed data $\mathbf{y} \in \mathbb{R}^N$. PS~\cite{eilers2003perfect} using second derivatives solves the smoothness regularized least-squares minimization problem as
\begin{align}
     \mathbf{x}^\ast &= \argmin_{\mathbf{x}} \frac{1}{2} \| \mathbf{x} - \mathbf{y} \|^2 + \frac{\lambda}{2} \| \mathbf{D} \mathbf{x} \|^2 \\ \nonumber
     &= \argmin_{\mathbf{x}} \sum_i \frac{1}{2} \left(x_i - y_i \right)^2 + \sum_{i'} \frac{\lambda}{2} \left( \mathbf{D} \mathbf{x} \right)_{i'}^2,
\end{align}
which has a closed-form solution that filters the noisy data as $\mathbf{x}^\ast = \left( \mathbf{I} + \lambda \mathbf{D}^{\top} \mathbf{D} \right)^{-1} \mathbf{y}$. In Equation 1, $\| \bullet \|$ is Euclidean norm. The matrix $\mathbf{I} \in \mathbb{R}^{N \times N}$ is the identity matrix and $\mathbf{D} \in \mathbb{R}^{N \times N-2}$ is the discrete second order derivative operator. We denote each data point of the second derivative as $i'$ that corresponds to the $i$-th data point. The parameter $\lambda \in \mathbb{R}_{\geq 0}$ balances the first data fidelity term and the second smoothness term. However, as Eilers suggested~\cite{eilers2003perfect}, $\lambda$ balances these terms only globally as fixed over all data points, but not locally at each point. To see this intuitively, we assume that $\lambda$ is given to balance the two terms $\frac{1}{2} \left(x_i^{\circ} - y_i \right)^2$ and $\frac{\lambda}{2} \left( \mathbf{D} \mathbf{x}^{\circ} \right)_{i'}^2$ to obtain $\mathbf{x}^{\ast} \approx \mathbf{x}^{\circ}$ at the $i$-th data point.
Consider another $j$-th data point where $\frac{\lambda}{2} \left( \mathbf{D} \mathbf{x}^{\circ} \right)_{j'}^2 \gg \frac{\lambda}{2} \left( \mathbf{D} \mathbf{x}^{\circ} \right)_{i'}^2$, and then, $\frac{\lambda}{2} \left( \mathbf{D} \mathbf{x}^{\ast} \right)_{j'}^2$ would be underestimated allowing a considerably larger deviation from the observation, which causes peak suppression.

\subsection{Locally self-adjustive penalized smoothing (LSA-PS)}
We consider locally balancing the smoothness against this unbalanced smoothing of PS that causes peak suppression. To this end, we redefine the cost function as $\sum_i \frac{1}{2} \kappa_i^2 \left(x_i - y_i \right)^2 + \sum_{i'} \frac{\lambda}{2} \left( \mathbf{D} \mathbf{x} \right)_{i'}^2$ and discuss how we give $\left\{ \kappa_i \right\}_{1 \leq i \leq N}$ for locally balancing the data fidelity and smoothness besides the parameter $\lambda$ that determines the global smoothness.  

We assume that the noise level is uniform over the data points; however we require no statistical assumption explicitly: If necessary, data are rescaled under an inhomogeneous noise level. Then, data fidelity and smoothness terms are balanced over all data points if, at the $i$-th data point, $\kappa_i$ equals to the second derivative of the smoothed data. Finding such $\left\{ \kappa_i \right\}_{1 \leq i \leq N}$ and thus the smoothed data simultaneously is a self-consistent problem that imposes a nonconvex problem that depends on initialization and intricate tuning for algorithm convergence. To obtain a unique solution by a single filtering, we approximate the second derivatives as locally as possible while avoiding over-fitting to data, and then we use them as $\left\{ \kappa_i \right\}_{1 \leq i \leq N}$. For that, we use the second-order polynomial fitting of selected data points around each data point. We consider five data points around each data point for regression, because the second-order polynomial overfits to three data points as it has three coefficients, and we use five data points for regression as the minimal local choice to avoid over-fitting. More data points can be used and incorporated in automatic parameter selection; however, it would less appropriately reflect the local characterization of data.

\begin{algorithm}[t]
\caption{LSA-PS}\label{algo1}
\begin{algorithmic}[1]
\Require noisy data $\mathbf{y} \in \mathbb{R}^{N}$, (scaled) global smoothness $\bar{\lambda}>0$, option "clip" $\in \left\{ {\rm on}, {\rm off} \right\}$ to avoid over-smoothing
\Ensure{smoothed data ${\mathbf{x}^\ast}$}
\State $\mathbf{A} = {\rm diag} \left( \left\{ {(2a_i^{\ast})^2} \right\}_{1 \leq i \leq N} \right)$ by solving Equation 2
\State $\lambda = \bar{\lambda} \cdot {\rm median}(\mathbf{A})$
\If{clip = {\rm on}}
\State $\mathbf{A} = {\rm diag} \left( \left\{ {\rm min} \left( (2a_i^{\ast})^2, {\rm median}(\mathbf{A}) \right) \right\}_{1 \leq i \leq N} \right)$
\EndIf
\State $\mathbf{x}^\ast = \left( \mathbf{A} + \lambda \mathbf{D}^{\top} \mathbf{D} \right)^{-1} \mathbf{A} \mathbf{y}$ (or use Cholesky decomposition)
\end{algorithmic}
\end{algorithm}

\begin{algorithm}
\caption{Parameter selection for LSA-PS}\label{algo2}
\begin{algorithmic}[1]
\Require $\mathbf{y} \in \mathbb{R}^{N}$, "clip" $\in \left\{ {\rm on}, {\rm off} \right\}$, parameter set $\Lambda$
\Ensure{optimal parameter $\bar{\lambda}^\ast$}
\For {$\bar{\lambda} \in \Lambda$}
    \State $\mathbf{H}(\bar{\lambda}) = \left( \mathbf{A} + \lambda \mathbf{D}^{\top} \mathbf{D} \right)^{-1} \mathbf{A}$ ($\bar{\lambda}$ to $\lambda$ in Algorithm 1)
    \State $\mathbf{x}(\bar{\lambda}) = \mathbf{H}(\bar{\lambda}) \mathbf{y}$
    \State $\mathbf{r}(\bar{\lambda})$: $r_i(\bar{\lambda}) = (y_i - x_i(\bar{\lambda})) ./ (1 - H_{ii}(\bar{\lambda}))$
    \State $S_{\rm cv}(\bar{\lambda}) = \sqrt{\mathbf{r}(\bar{\lambda})^\top \mathbf{A}^{-1} \mathbf{r}(\bar{\lambda})/N}$
\EndFor
\State $\bar{\lambda}^{\ast} = \argmin_{\bar{\lambda} \in \Lambda} \left\{ S_{\rm cv}(\bar{\lambda}) \right\}_{\bar{\lambda} \in \Lambda}$
\end{algorithmic}
\end{algorithm}

Second-order polynomial regression using five data points around and including the $i$-th data point is
\begin{align}
\left\{ {a^{\ast}_i, b^{\ast}_i, c^{\ast}_i} \right\} = \argmin_{a_i, b_i, c_i} \sum_{w=1}^{5} \left( y_{i,w} - \left(a_i \xi_{i,w}^2 + b_i \xi_{i,w} + c_i \right) \right)^2,
\end{align}
where we used a local coordinate $\mathbold{\xi}_i = \left(-2, -1, 0, 1, 2 \right)^{\top}$ at the $i$-th data point (located at 0), the $w$-th element of which was written as $\xi_{i, w}$, and $y_{i,w}$ represents the observed value. Equation 2 has a unique solution, and we have the second-order coefficients $\left\{{a^{\ast}_i}\right\}_{1 \leq i \leq N}$; then we set $\kappa_i^2 = (2a_i^{\ast})^2$ to have
\begin{align}
     \mathbf{x}^\ast &= \nonumber \argmin_{ \mathbf{x}} \sum_i \frac{(2a_i^{\ast})^2}{2} \left(x_i - y_i \right)^2 + \sum_{i'} \frac{\lambda}{2} \left(\mathbf{D} \mathbf{x}\right)_{i'}^2 \\ \nonumber
     &= \argmin_{\mathbf{x}} \frac{1}{2} \left( \mathbf{x} - \mathbf{y} \right) \mathbf{A} \left( \mathbf{x} - \mathbf{y} \right) + \frac{\lambda}{2} \| \mathbf{D} \mathbf{x} \|^2 \\
     &= \left( \mathbf{A} + \lambda \mathbf{D}^{\top} \mathbf{D} \right)^{-1} \mathbf{A} \mathbf{y},
\end{align}
where $\lambda \in \mathbb{R}_{\geq 0}$ is the only parameter that determines the global smoothness, and the matrix $\mathbf{A} \in \mathbb{R}^{N \times N}$ is a diagonal matrix with its $(i,i)$-th element as ${(2a_i^{\ast}})^2$, which is obtained from the data itself. Equation 3 is efficiently computed using Cholesky decomposition as Eilers implemented~\cite{eilers2003perfect}. Equation 3 is the same as the Elier's formulation when we replace $\mathbf{A}$ with the weight matrix in PS. The role of $\mathbf{A}$ is different however, as the weights with binary values of $\left\{ 0, 1 \right\}$ are considered for self-cross validation, or the weights are used for baseline correction~\cite{eilers2005baseline}.

We control the global smoothness $\lambda$ through a scaled one, $\bar{\lambda}$, in a data-driven manner as $\lambda = \bar{\lambda} \cdot {\rm median}(\mathbf{A})$ to use the same parameter scale for PS and the proposed method. This provides a practical convenience but is not the essence of the algorithm.

Compared to PS, the proposed method can smooth data more strongly while preserving sharp peaks. However, it tends to overly smooth data points near the beginning and at the end of peaks, where the second-order polynomial does not approximate data well. Then, we introduce a simple strategy to clip the elements of $\mathbf{A}$, which approximate the local smoothness, with ${\rm median}(\mathbf{A})$: This does not require additional parameters, but the clipping value could optionally be controlled as a parameter.

The proposed locally self-adjustive penalized smoother (LSA-PS, so named after Eilers' PS) is shown in Algorithm 1.

\subsection{Parameter selection}
Eilers adopted the leave-one-out cross validation (CV)~\cite{hastie1990generalized} for parameter selection in PS. Although peak suppression is mitigated in LSA-PS, selected parameters were underestimated for both PS and LSA-PS. Eilers suggested that even such a small smoothness could leave the estimated noise uncorrelated, which is a statistical assumption of the CV, and correlated local behaviour could be the source of the underestimation. We infer that small bumps and dimples occasionally caused by noise prohibit sufficient smoothing. Then, we empirically modified the CV loss (error) $S_{\rm cv}$ as in Algorithm 2: The original loss~\cite{eilers2003perfect} is $\sqrt{ \mathbf{r}^\top \mathbf{r} / N }$, whereas the modified loss is $\sqrt{ \mathbf{r}^\top \mathbf{A}^{-1} \mathbf{r} / N }$ in a data-driven manner for improving underestimation by less accounting such bumps and dimples from noise. Our modification is only empirical and lacks theory; however, it achieved nice performance for all simulation and real experiments in this study. CV is an option for series with randomness such as spectroscopic data; however, a comparison with other criteria of parameter selection~\cite{harville1974bayesian, akaike1973information, akaike1998information} would also be valuable.

\section{Experimental results and discussion}

We evaluated LSA-PS using simulation and real data, and compared the results against those of PS, SG, and Gaussian smoothing. We used MATLAB 2021b~\cite{MATLAB} for all implementations and evaluations, and we used its implementations of SG and Gaussian smoothing. A quantitative analysis was conducted using the simulation (Section 3.1, Figure 1). We qualitatively show how LSA-PS allows sufficient smoothing while preserving sharp peaks (Figure 2), as well as an improved automated smoothing (Figure 3, 4). We used Raman spectra in the real experiment to demonstrate automated smoothing and peak detection (Section 3.2, Figure 5-8). In the supplementary material, we present the results for X-ray diffraction (XRD)~\cite{sun2018accelerating, oviedo2019fast} and Fourier-transform Infrared Spectroscopy (FTIR)~\cite{wang2023influence, wang2023IRdataset} data.

\begin{figure*}[t]
\centering
\includegraphics[width=16cm]{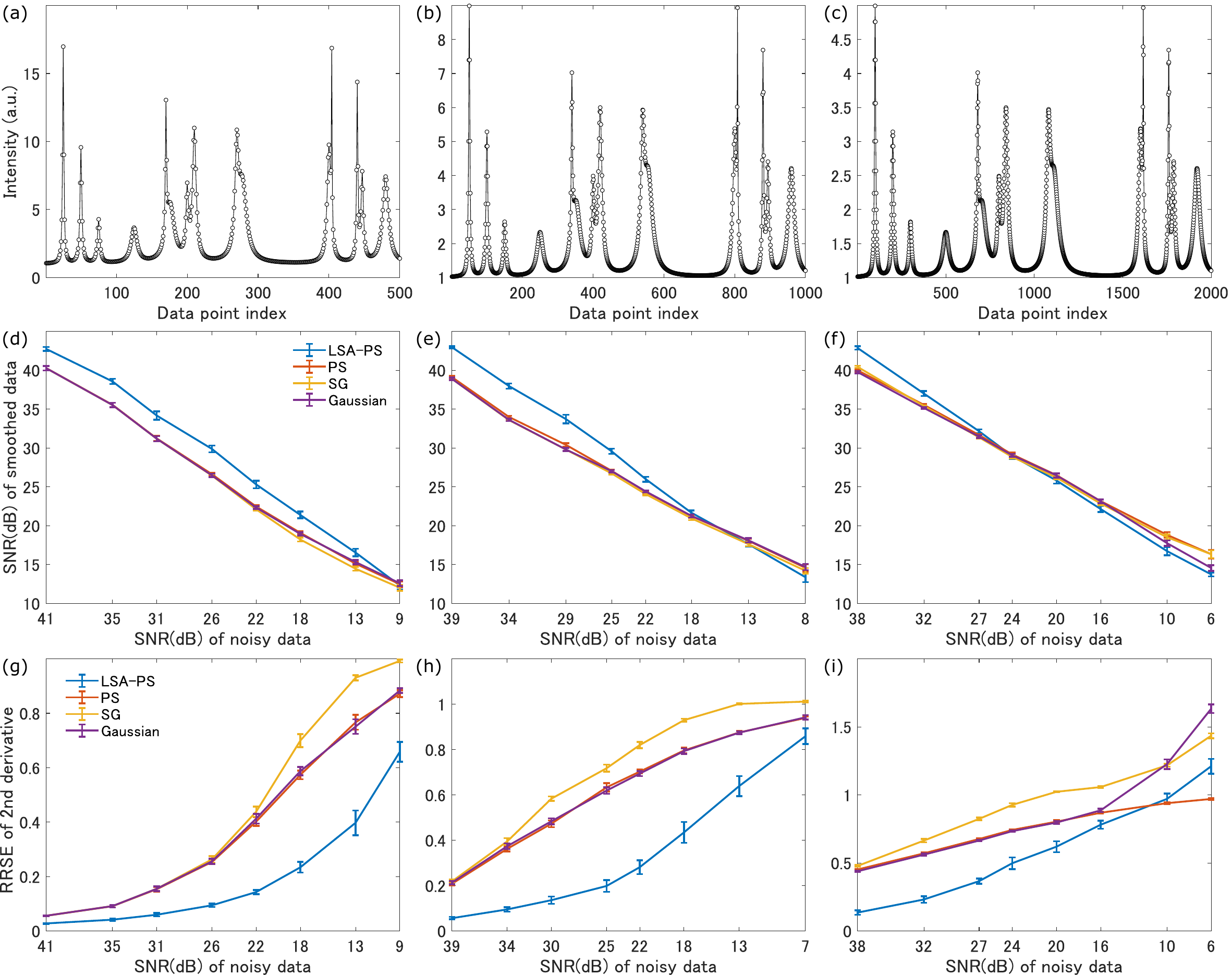}
\vspace{-0mm}
\caption{\small{Top: simulated data with (a) 500 data points, (b) 1000 data points, and (c) 2000 data points. Middle: averages and standard deviations of the SNRs of smoothed data using LSA-PS, PS, SG, and Gaussian smoothing, against those of noise added data; (d-f) results obtained using the simulated data (a-c) in this order. Bottom: averages and standard deviations of the RRSEs of the second derivatives of the smoothed data against those of the noise added data; (g-i) results obtained using the simulated data (a-c).}}
\vspace{-0mm}
\end{figure*}

\begin{figure*}
\centering
\includegraphics[width=12cm]{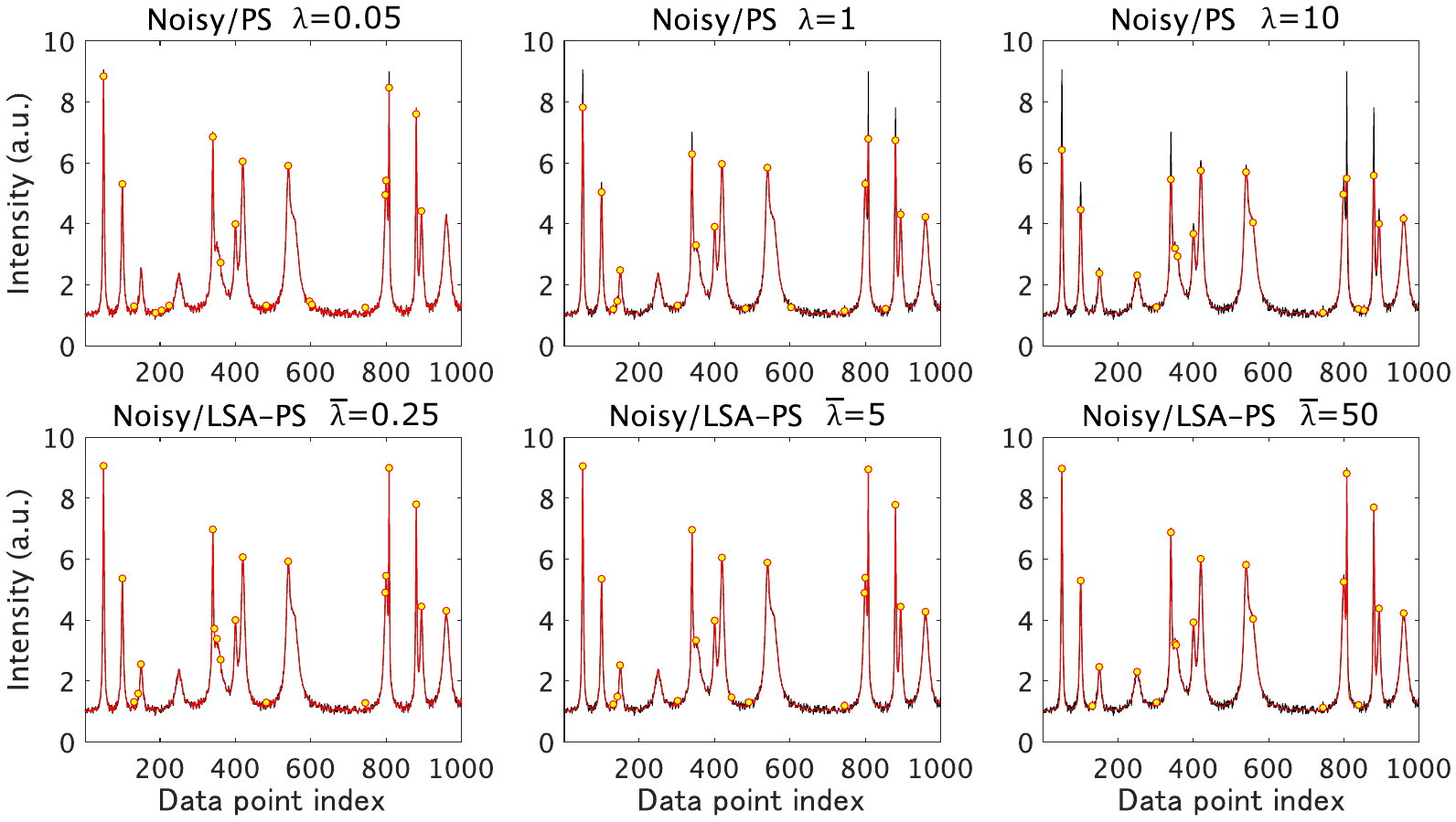}
\vspace{-1mm}
\caption{\small{Comparison of PS (top) and LSA-PS (bottom) with varying their smoothness parameters (increased from left to right).}}
\vspace{-0mm}
\end{figure*}

\begin{figure}
\includegraphics[width=8.75cm]{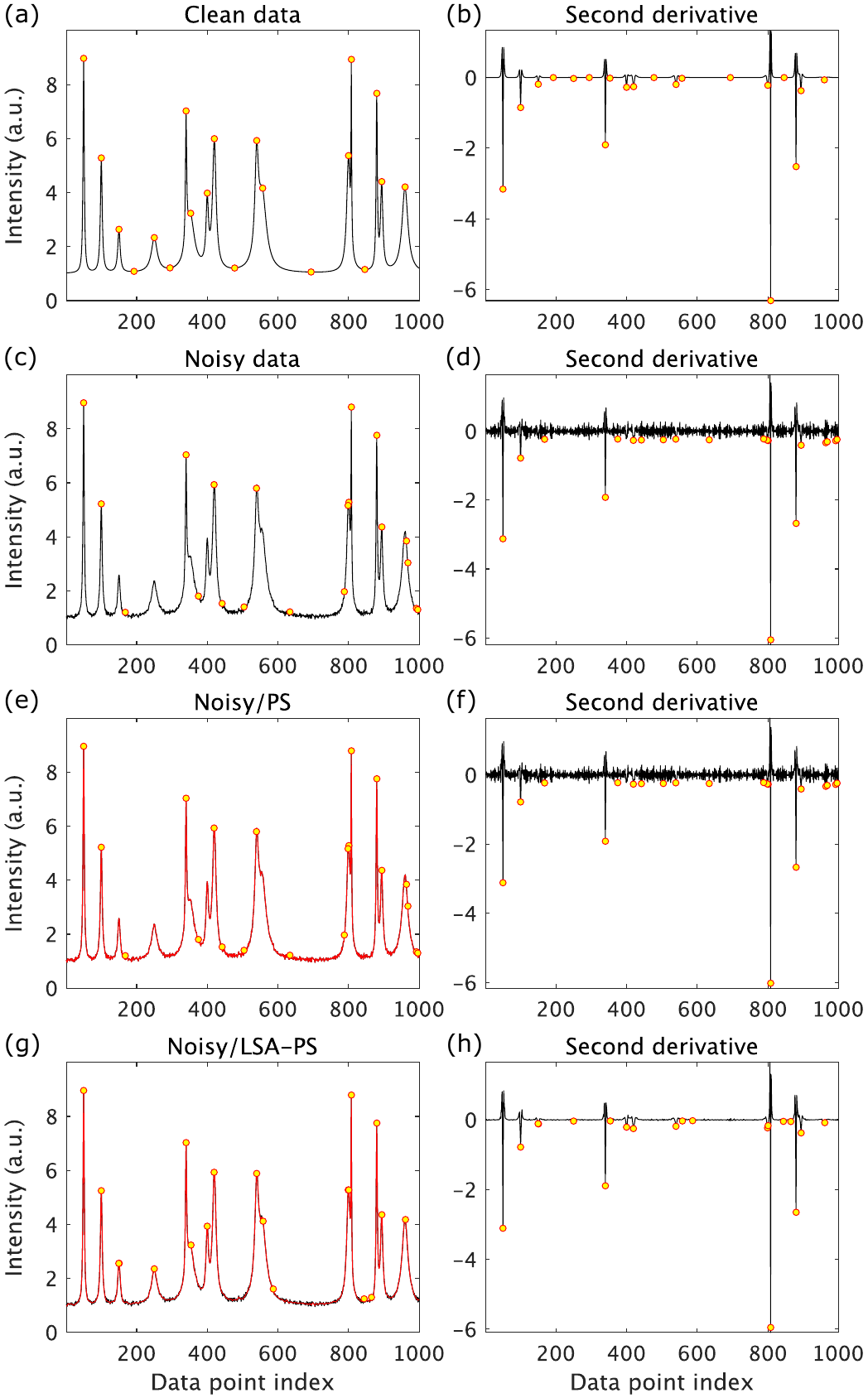}
\vspace{-3mm}
\caption{\small{(a) Simulated data with 1000 data points, (b) noisy data of 34 dB, results of (c) PS and (d) LSA-PS, with their second derivatives on the right and 20 detected peaks.}}
\vspace{-0mm}
\end{figure}

\begin{figure}
\includegraphics[width=8.75cm]{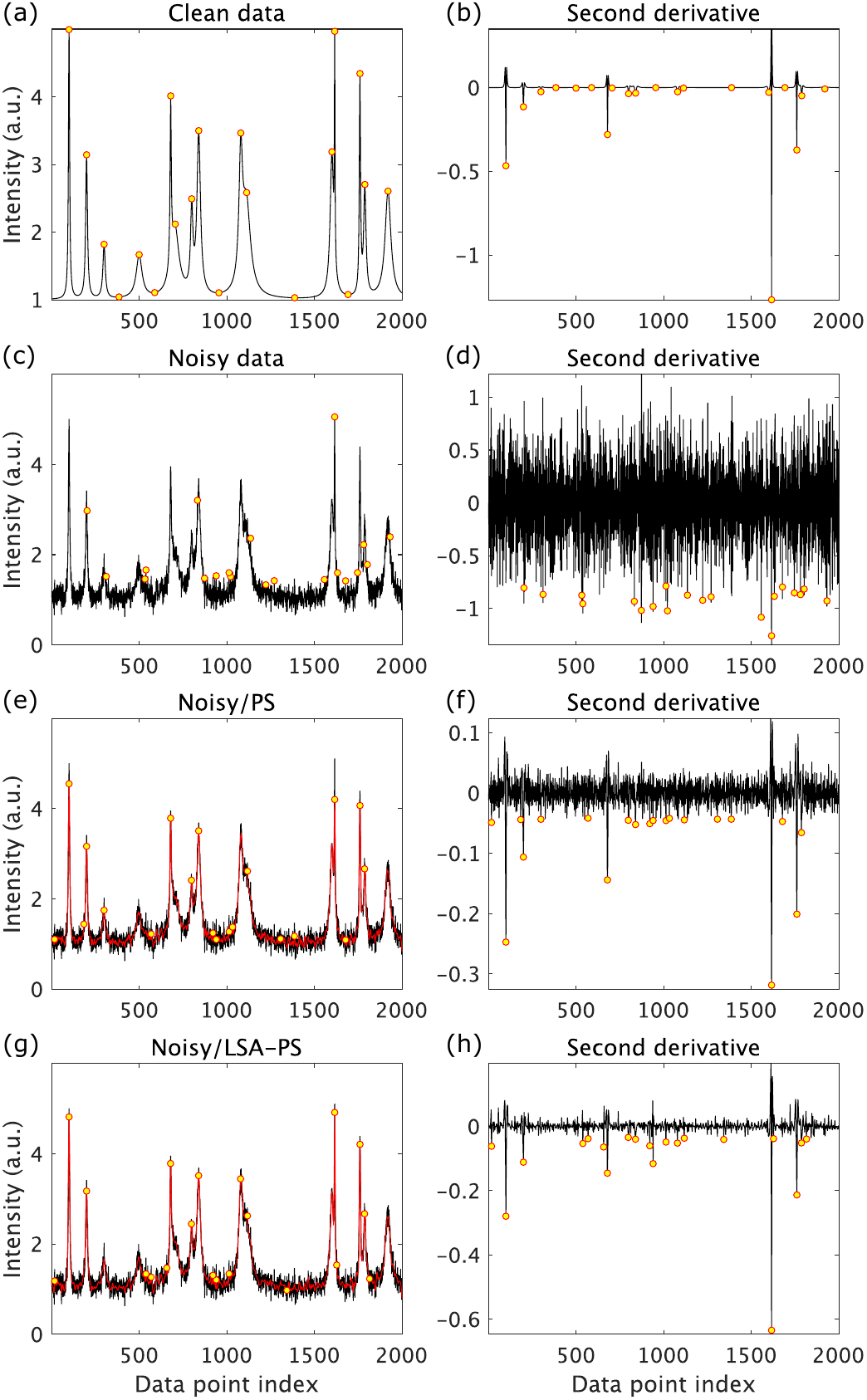}
\vspace{-3mm}
\caption{\small{(a) Simulated data with 2000 data points, (b) noisy data of 21 dB, results of (c) PS and (d) LSA-PS, with their second derivatives on the right and 20 detected peaks.}}
\vspace{-0mm}
\end{figure}

\subsection{Simulation experiment}
We simulated data $\mathbf{x}^{\circ}$ of the superposition of 15 Lorentzian functions with different peak positions, heights, and widths. Two of them have close peak positions. Besides the 15 peaks, there also exist weak peaks caused by the addition of two neighboring Lorentzian functions. These peaks can be detected using the second derivative without noise; however, they were difficult to detect under noise. The resolution is an important factor for peak suppression after smoothing as Yao et al.~\cite{yao2021yield} suggested, as well as peak detection using second derivative. For investigating that effect, we considered three different resolutions for the same simulated data with taking 500 (Figure 1 (a)), 1000 ($\times 2$, Figure 1 (b)) and 2000 ($\times 4$ resolution, (c)) data points.

We added Gaussian noise $\bm{\varepsilon}$ with its mean of zero and various standard deviations, and we smoothed the noisy data $\mathbf{y} = \mathbf{x}^\circ + \bm{\varepsilon}$ using each method. The parameters of each method ($\bar{\lambda}$ for LSA-PS, $\lambda$ for PS, and the window size ("framelen" in MATLAB's implementation) and the degree of polynomials for SG, and the window size for Gaussian smoothing in the supplementary material, Table 1) were varied, and we selected the best parameters in terms of SNRs $10 \log_{10} \left( \| \mathbf{x}^\circ \|^2 / \| \bm{\varepsilon} \|^2 \right)$, to evaluate the potential performances of these methods. Figures 1 (d-f: resolutions correspond to (a-c)) plots the SNRs (ten-trials' average and standard deviation, which may not be the best way for statistically measuring the distribution of the SNRs) of the smoothed data against those of the noisy data.

For the data with 500 points, LSA-PS denoised the best when the SNR of the noisy data is above about 9 dB. PS and Gaussian smoothing have close SNRs after denoising, and SG showed the lowest SNRs. For the data with 1000 and 2000 points, LSA-PS performed better than the other methods against moderate SNRs, but became the worst when the SNR of the noisy data decreased. We consider that convolution-based smoothing is incapable of reducing noise of such highly noisy data. PS and Gaussian smoothing are comparable, but PS was better. Comparing the SNRs after smoothing at the same SNRs of the noisy data revealed that improvement using LSA-PS was more distinct when the resolution (e.g. wavelength step in spectroscopic measurement) is lower. We consider the reason that LSA-PS could find local shapes of data but they might come from noise when the resolution is quite high, especially with small peaks and without peaks.

Figures 1 (g-i) show the plots of root relative squared error (RRSE) $\| \mathbf{D} \mathbf{x}^\ast - \mathbf{D} \mathbf{x}^\circ \| / \| \mathbf{D} \mathbf{x}^\circ \|$ against the SNRs of the noisy data. LSA-PS was better than the other methods for the data with 500 and 1000 points. With 2000 data points, LSA-PS also outperformed the other methods for most cases, but PS was better for data with two smallest SNRs. Second derivatives are closely related to peak detection and the sharpness of peaks, and therefore, these results suggest that LSA-PS could better detect peaks while retaining sharp peaks. 

Figure 2 shows how LSA-PS avoids suppression of sharp peaks in comparison to PS. We show 20 peaks, with five margins in addition to the Lorentzian 15 peaks, detected by counting the 20 top sharpness values evaluated with the absolute values of second derivatives. We increased the parameters from left to right as their smoothness in the same columns are visually comparable at regions without peaks. Noisy data were overlaid with the smoothed data in each plot. At the rightmost column of Figure 2, both PS and LSA-PS finally detected the prominent 15 peaks; whereas PS suppressed the sharp peaks, and LSA-PS retained them. 

In Figure 1, PS and Gaussian smoothing were comparable for most cases. However, we can use parameter selection for PS. Next, we discuss the automatic application of PS and LSA-PS using the CV for PS and Algorithm 2 for LSA-PS. The parameters used for the CVs ($\Lambda$ for PS and Algorithm 2; the supplementary material, Table 2) were fixed throughout the simulation and in the real experiments.

We show two examples of smoothed data and the peak detection results obtained using the simulated data of 1000 points. The first example is of a high SNR with 20 peak detection. Figure 3 (a) is without noise; (c) is a noisy data of 34 dB, where 6 peaks were missed; (e) is the smoothed data and detected peaks, obtained using PS and its automatically selected parameter; however, the smoothing was so weak that the peaks were still missed. In contrast, all peaks were detected when using LSA-PS (g). Comparing the second derivatives revealed that LSA-PS reduced noise better than that using PS. The second example is of a low SNR of 21 dB. Figure 4 (a) shows the the clean data with 20 peaks. For both data smoothed by PS and LSA-PS, we detected 10 peaks, some of which were at different positions. Although peak detection performance was comparable for such highly noisy data, LSA-PS smoothed more strongly while retaining the sharp peaks compared to PS. The peak detection results evaluated in this study are based only on second derivative, and thus, the peak intensities were not considered. Using peak intensities in addition to the second derivatives would help to detect peaks, but this optional consideration is beyond the scope of this study.

An additional result using a simulation data with background is shown in the supplementary material, Figure 12, which demonstrates that LSA-PS can be used for data with non-flat baselines.

\begin{figure}[t]
\includegraphics[width=8.75cm]{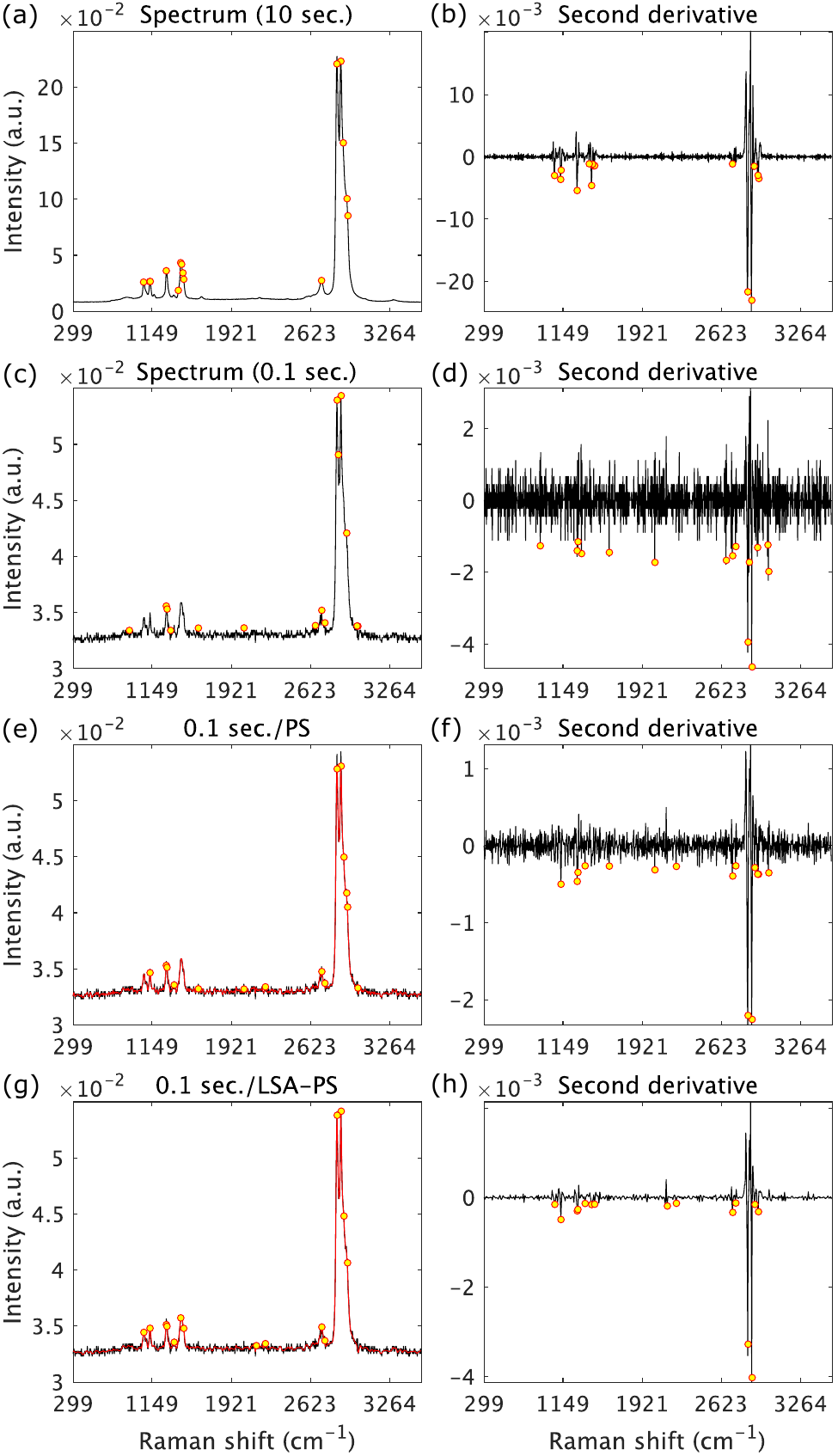}
\vspace{-3mm}
\caption{\small{Raman spectra of PE with (a) 10 and (c) 0.1 second exposures, smoothed spectra using (e) PS and (g) LSA-PS; second derivatives on the right.}}
\vspace{-0mm}
\end{figure}

\begin{figure}
\includegraphics[width=8.75cm]{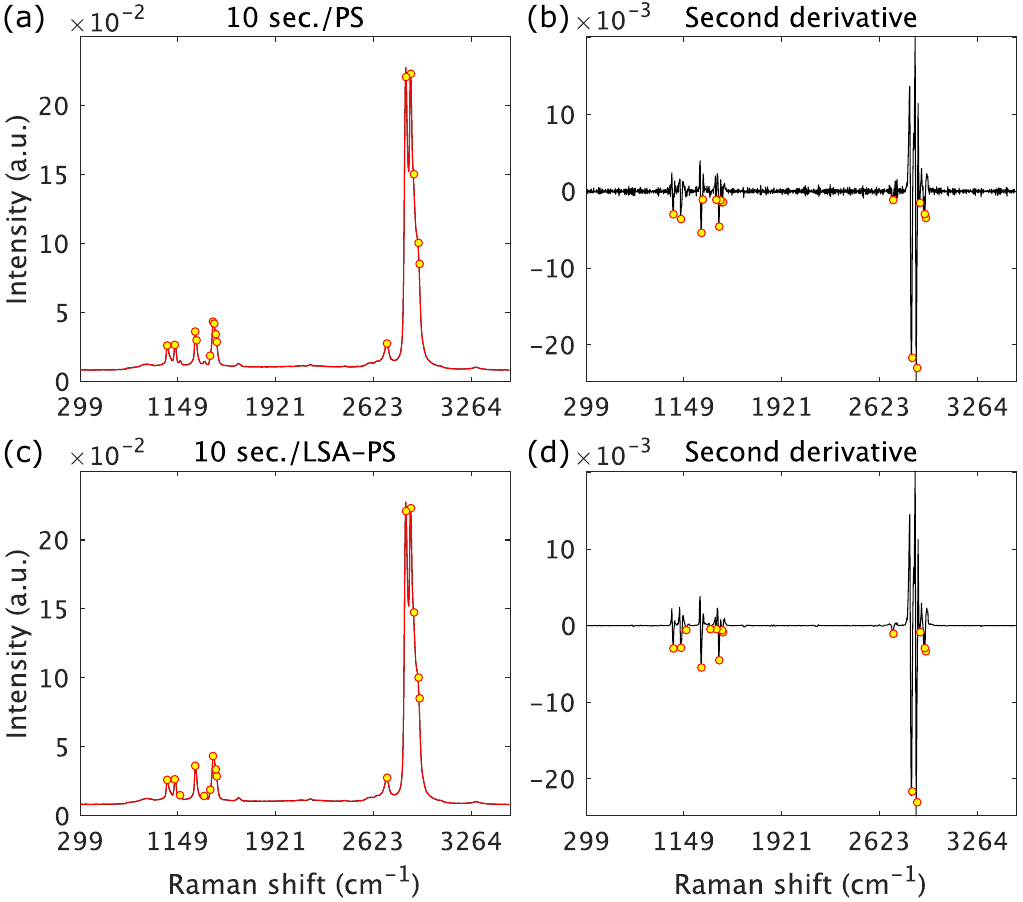}
\vspace{-3mm}
\caption{\small{Smoothed spectra of 10 second exposed spectrum of PE using (a) PS and (b) LSA-PS; second derivatives on the right.}}
\vspace{-0mm}
\end{figure}

\subsection{Real experiment}
For evaluation in practical scenarios, we applied PS and LSA-PS with their parameter selection criteria.

We obtained Raman spectra of a commercial product polyethylene (PE) film using a laser Raman spectrometer NRS-5500 (JASCO Corporation). We measured the sample with two exposure times of 10 and 0.1 seconds, by fixing the excitation wavelength of 532 nm, the intensity of 3 mW, and by performing a single measurement. The aperture diameter was 4000 $\mu m$. Figures 5 (a) and (c) show the Raman spectra obtained at exposure times of 10 and 0.1 seconds, respectively, and their second derivatives are shown in Figures 5 (b) and (d).

We applied PS and LSA-PS for the noisy spectrum (Figure 5 (c)). The CVE plots are shown in the supplementary material, Figure 10. The smoothed spectra and detected 15 peaks are shown in Figures 5 (e) for PS and (g) for LSA-PS, as well as their second derivatives in (f) and (h). The larger smoothness for LSA-PS resulted in a more smoothed second derivative (h) compared to PS (f). LSA-PS smoothed the noisy spectrum more strongly, while retaining two sharp peaks at 2844 and 2878 ${\rm cm}^{-1}$, which correspond to asymmetric and symmetric ${\rm CH_2}$ stretches. LSA-PS also detected peaks that could be attributed~\cite{nava2021raman, huang1971raman} (with a direct reference of the literature~\cite{nava2021raman}) to the asymmetric and symmetric ${\rm C-C}$ stretches (1063 and 1124 ${\rm cm}^{-1}$), ${\rm CH_2}$ twist (1293 ${\rm cm}^{-1}$), ${\rm CH_2}$ vibration (1436 ${\rm cm}^{-1}$), the overtone from ${\rm CH_2}$ (2717 ${\rm cm}^{-1}$), and the Fermi resonance (2931 ${\rm cm}^{-1}$). On the other hand, PS failed to detect two of them from  asymmetric ${\rm C-C}$ stretch (1063 ${\rm cm}^{-1}$) and ${\rm CH_2}$ vibration (1436 ${\rm cm}^{-1}$). We consider that the small differences of peak positions from that reported in the literature, which does not directly affect our evaluation, may be attributed to the difference in the wavelength resolution or wavelength collection accuracy. Some of the other peaks may be from remaining noise.

In Figure 6, we show the results of PS and LSA-PS applied for the relatively clean spectrum with 10 second exposure and 15 detected peaks. The CVE plots are shown in the supplementary material, Figure 11. Both PS and LSA-PS detected characteristic peaks of PE including ${\rm CH_2}$ wagging, and only LSA-PS detected two small peaks from ${\rm CH_2}$ bend (1169 ${\rm cm}^{-1}$) and the peak at 1365 ${\rm cm}^{-1}$, which may be from CH bend: These weak peaks were enhanced in the second derivative for the LSA-PS applied spectrum (Figure 6 (d)).

The computational speeds were compared (in the supplementary material, Figure 9). For a single call and run, PS was the fastest, followed by LSA-PS, and SG. For the automated application using the CVs, PS was faster than LSA-PS, but they were relatively comparable in contrast to the single call, because local smoothness is computed once and the other computational complexities are comparable.

\begin{figure}[t]
\includegraphics[width=8.75cm]{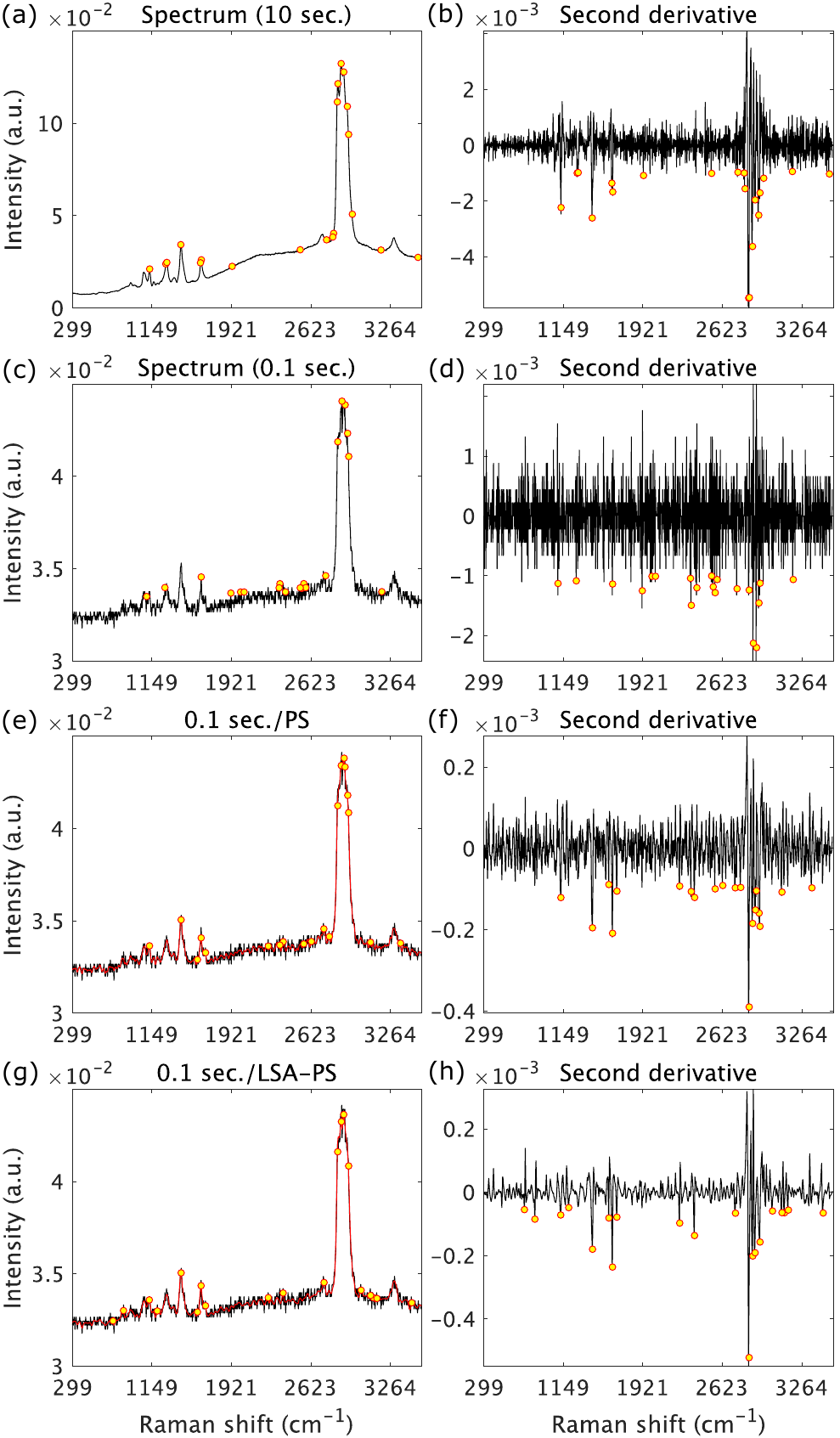}
\vspace{-3mm}
\caption{\small{Raman spectra of Nylon with (a) 10 and (c) 0.1 second exposures, smoothed spectra using (e) PS and (g) LSA-PS; second derivatives on the right.}}
\vspace{-0mm}
\end{figure}

\begin{figure}[t]
\includegraphics[width=8.75cm]{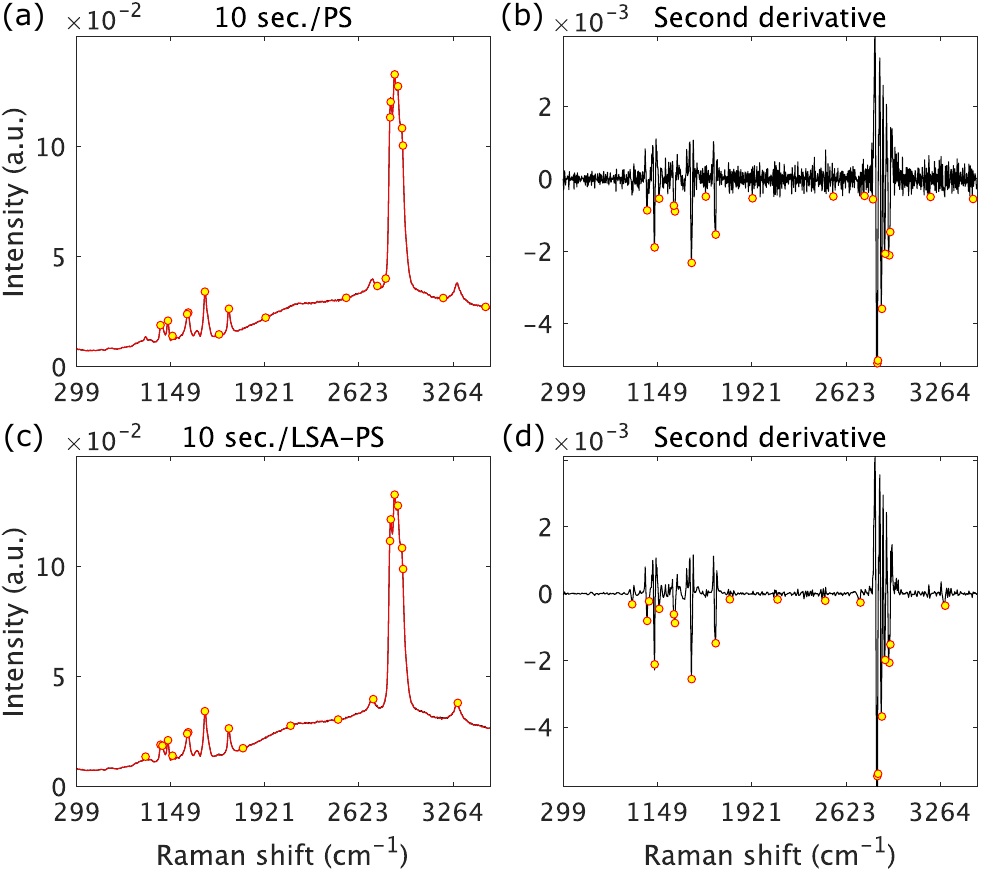}
\vspace{-3mm}
\caption{\small{Smoothed spectra of 10 second exposed spectrum of Nylon using (a) PS and (b) LSA-PS; second derivatives on the right.}}
\vspace{-0mm}
\end{figure}

We provide another example using a Nylon film as a difficult scenario for bothe PS and LSA-PS. The measurement condition was the same for PE. Figures 7 (a) and (c) are spectra obtained at exposure times of 10 and 0.1 seconds with 20 detected peaks; (b) and (d) show their second derivatives. We observed a background in addition to noise of a relatively high level compared to the example of PE. Figures 7 (e) and (g) show the smoothed spectra using PS and LSA-PS; (f) and (h) are their second derivative spectra. Both PS and LSA-PS failed to detect some peaks. However, LSA-PS smoothed the data more strongly while retaining sharp peaks, and reducing the fluctuations in the second derivative spectra.

In Figure 7 (a), some important peaks are not detected, while some small fluctuations are detected as false peaks. We applied PS and LSA-PS for the spectrum and show the results in Figure 8. Although PS smoothed the spectrum to some extent, most peaks were still missed because of underestimated smoothness. In contrast, for the LSA-PS applied spectrum (Figure 8 (c)), we detected prominent peaks at 931, 1063, 1128, 1165, 1301, 1440, 1637, 2727, 2848, 2847, 2925, 2932, 3295 ${\rm cm}^{-1}$, most of which can be attributed to those originated from Nylon.

In the supplementary material, we also show additional results for XRD data~\cite{sun2018accelerating, oviedo2019fast} (Figure 13-17) and FTIR data~\cite{wang2023influence, wang2023IRdataset} (Figure 18-20). The results for the XRD data prominently demonstrate the improved noise reduction and peak detection performance of LSA-PS with retaining sharp peaks.

\subsection{Suitable scenario for using LSA-PS}
Throughout the simulation and real experiments, we considered that LSA-PS is useful in scenarios where noise level is moderate, wavelength (or diffraction angle, etc.) resolution is moderate, and sharp peaks exist. Even moderate noise on measurement data affects peak analysis, and so a certain preprocessing is often required. Although LSA-PS should be used with care because it can  introduce bias as the other smoothing methods do, the automated noise reduction using LSA-PS is a nice alternative in such typical scenarios. On the other hand, when the noise level is high and/or the resolution is considerably fine, LSA-PS would not work so nicely, because the local characteristics of peaks cannot be captured using only the second derivatives, which is the basis of PS and LSA-PS. Smoothing may be insufficient to enhance peaks when the noise level is high; when the resolution is fine, the second derivative alone may not be enough to detect the peaks.

\section{Conclusions}
We presented a smoothing method that balances smoothness in a locally self-adjustive manner. Despite the method requiring only one smoothness parameter for tuning, it retains sharp peaks while sufficient smoothing. Further, we demonstrated automated application by estimating the smoothness parameter using a modified cross validation error. The simulation and real experiments suggest that the proposed method could become a nice alternative for the existing smoothing methods, with comparable computational speed, against data with moderate noise level and moderate resolution. In such scenarios, the proposed method can be used as an instant option to automatically smooth data and find peak candidates. For data of relatively low SNRs and/or data with fine resolutions, the smoothed data and detected peak candidates should be treated more carefully.

\section*{Acknowledgement}
We would like to thank Felipe Oviedo, Tonio Buonassisi, and any contributor who shared the XRD data~\cite{sun2018accelerating, oviedo2019fast} and permission from the publisher (https://www.nature.com/). We would also like to thank Tianyuan Wang and any contributor who shared the FTIR data~\cite{wang2023influence, wang2023IRdataset} and permission from the publisher (https://www.sciencedirect.com/journal/data-in-brief).

\section*{Author Contributions}
Keisuke Ozawa conceived this study, formulated and implemented the method, conducted experiments, and wrote the manuscript. All authors contributed to the discussion, interpretation of experimental results, and reviewing the manuscript.

\normalsize
\bibliography{references}


\clearpage
\onecolumn
\section*{Supplementary results}
We provide some additional results that support the discussion in the main text and the possible users of LSA-PS.

\subsection*{Parameters}
Table 1 lists the parameters of the four smoothing methods used for the quantitative comparison in Figure 1. We selected the best SNR and RRSE over the parameters for each method.

Table 2 lists the parameters used for the CVs of PS and LSA-PS. Owing to the scaling from $\lambda$ to $\bar{\lambda}$ using the median of second derivatives (Section 2.2), we used the same parameters for both PS and LSA-PS.

\begin{table}[h]
\renewcommand*{\arraystretch}{1.1}
\centering
\caption{\small{Parameters used for the quantitative comparison in Figure 1.}}
\vspace{0 mm}
\scalebox{1}
{
\begin{tabular}{|l|l|}
\hline
 Method &\multicolumn{1}{c|}{Parameter} \\
\hline
PS & $\lambda \in \left\{0.0001, 0.001, 0.01, 0.1, 0.5, 1, 2, 3, 4, 5, 6, 7, 8, 9, 10, 20, 50, 100\right\}$ \\
LSA-PS & $\bar{\lambda} \in \left\{0.0001, 0.001, 0.01, 0.1, 0.5, 1, 2, 3, 4, 5, 6, 7, 8, 9, 10, 20, 50, 100\right\}$ \\
SG & framelen $\in \left\{1, 3, 5, 7, 9, 11, 13, 15, 17, 19, 21, 23, 25, 27, 29, 31, 33, 35\right\}$; $1 \leq$ order $<$ framelen\\
Gaussian & window $\in \left\{1, 2, 3, 4, 5, 6, 7, 8, 9, 10\right\}$\\
\hline
\end{tabular}
}
\label{table:comparison}
\end{table}

\begin{table}[h]
\renewcommand*{\arraystretch}{1.1}
\centering
\caption{\small{Parameters used for the CV-based parameter selection ($\bar{\lambda} \in \Lambda$ in Algorithm 2).}}
\vspace{0 mm}
\scalebox{1}
{
\begin{tabular}{|l|l|}
\hline
 Method &\multicolumn{1}{c|}{Parameter} \\
\hline
PS & $\lambda \in \left\{0.001, 0.01, 0.1, 0.5, 1, 2.5, 5, 10, 25, 50, 100\right\}$ \\
LSA-PS & $\bar{\lambda} \in \left\{0.001, 0.01, 0.1, 0.5, 1, 2.5, 5, 10, 25, 50, 100\right\}$ \\
\hline
\end{tabular}
}
\label{table:comparison}
\end{table}

\subsection*{Computational speed}
Figure 9 shows the comparison of the computational times against the PE Raman spectrum of 870 data points. Figure 9 left compares the single call and execution of each method; PS was the fastest, and LSA-PS was the second, and SG was the third. These results depend on the implementations (Matlab2021b for SG and Gaussian smoothing), and this could be improved: For example, the calculation of local smoothness in LSA-PS can be parallelized. The speed of SG and Gaussian also may have rooms of improvement in practical usage. 

Figure 9 right compares the computational times of the automatic applications and shows that PS and LSA-PS are more comparable. Although the gap in speed between PS and LSA-PS comes from the computation of local smoothness, it is called once in the CVE calculation and thus, the total times are more comparable than the comparison of a single call and execution.

\begin{figure*}[h]
\centering
\includegraphics[width=14cm]{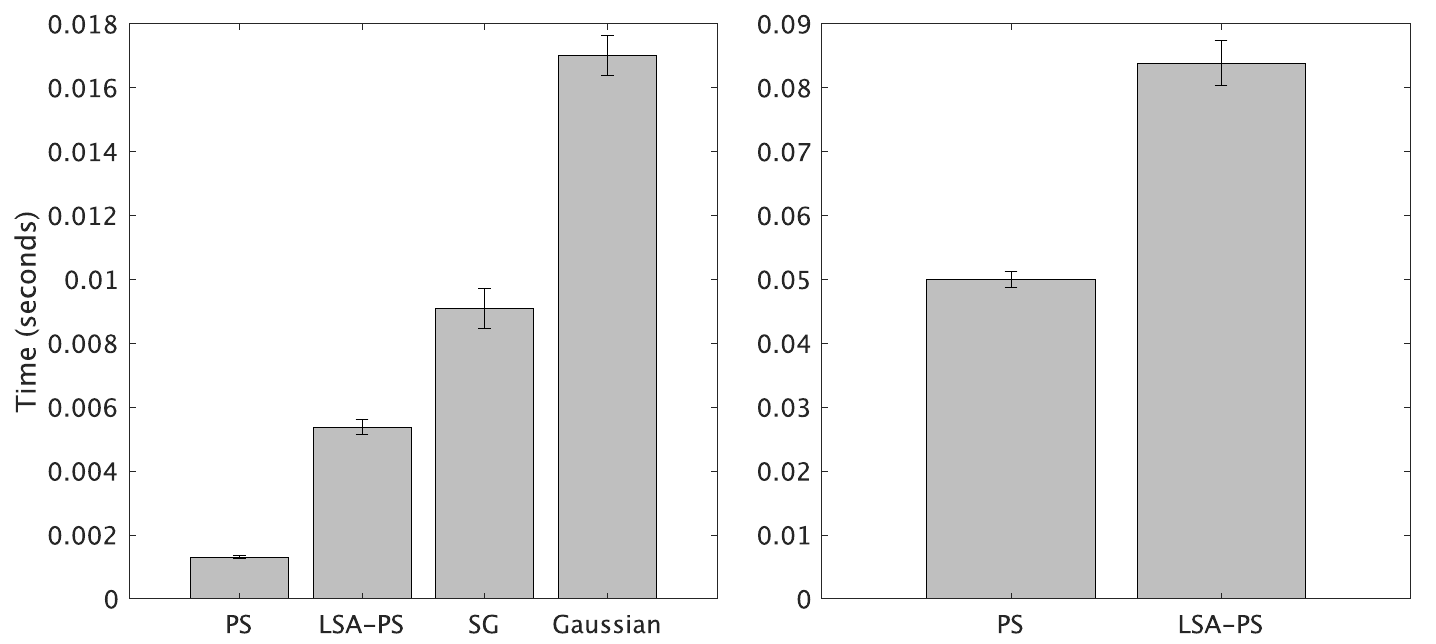}
\vspace{-0mm}
\caption{\small{CVE plots of (a) PS and (b) LSA-PS using Algorithms 1 and 2 for the Raman spectrum of PE with 0.1 second exposure.}}
\vspace{-0mm}
\end{figure*}

\clearpage
\subsection*{CVE plots of PS and LSA-PS}
Figures 10 and 11 show the CVE plots using PS and LSA-PS. We adjusted the scales to some extent by dividing the CVE for LSA-PS by the norm of $\mathbf{A}$ in Algorithm 2.

\begin{figure*}[h]
\centering
\includegraphics[width=11cm]{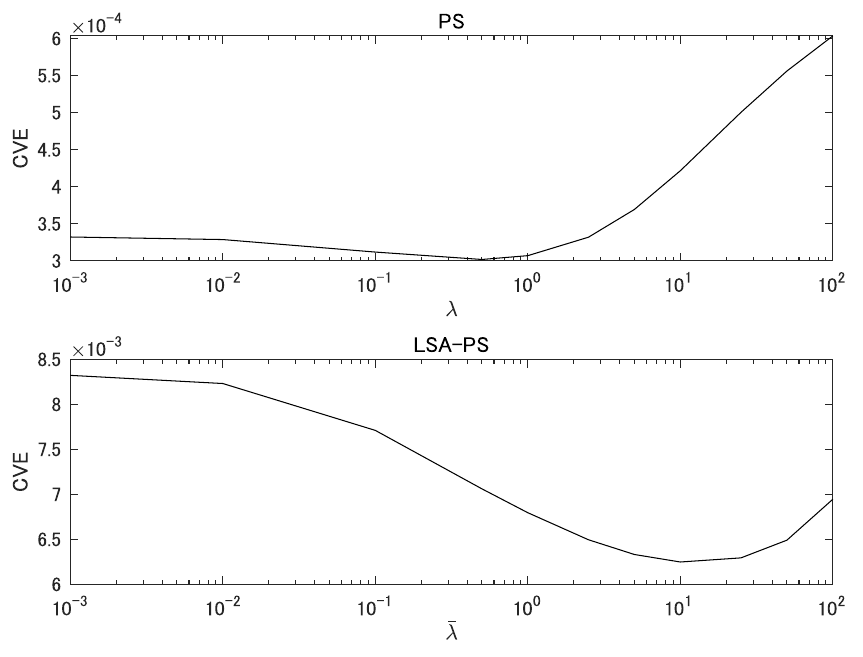}
\vspace{-0mm}
\caption{\small{CVE plots of (a) PS and (b) LSA-PS using Algorithms 1 and 2 for the Raman spectrum of PE with 0.1 second exposure.}}
\vspace{-4mm}
\end{figure*}

\begin{figure*}[h]
\centering
\includegraphics[width=11cm]{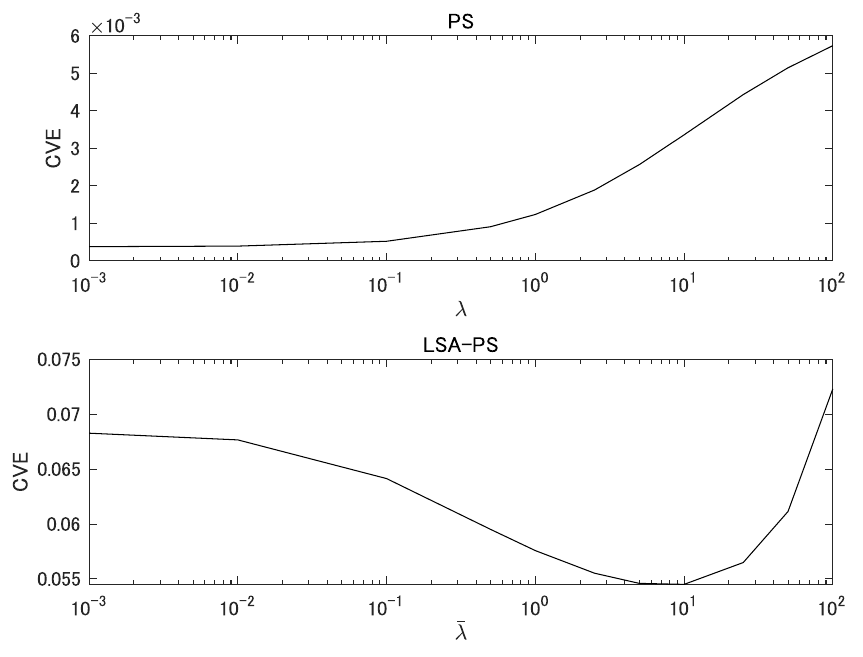}
\vspace{-0mm}
\caption{\small{CVE plots of (a) PS and (b) LSA-PS using Algorithms 1 and 2 for the Raman spectrum of PE with 10 second exposure.}}
\vspace{-0mm}
\end{figure*}

\clearpage
\subsection*{Comparison of PS and LSA-PS against a simulation data with background}
We have shown that LSA-PS performed nicely even against data with background (Figures 7 and 8 of Nylon Raman spectra and the following XRD data). We additionally show a simulation result with background data in Figure 12. LSA-PS better smoothed the data than PS with detecting more number of peaks, while retaining the sharp peak at 808 of data point index.

\begin{figure*}[h]
\centering
\includegraphics[width=11cm]{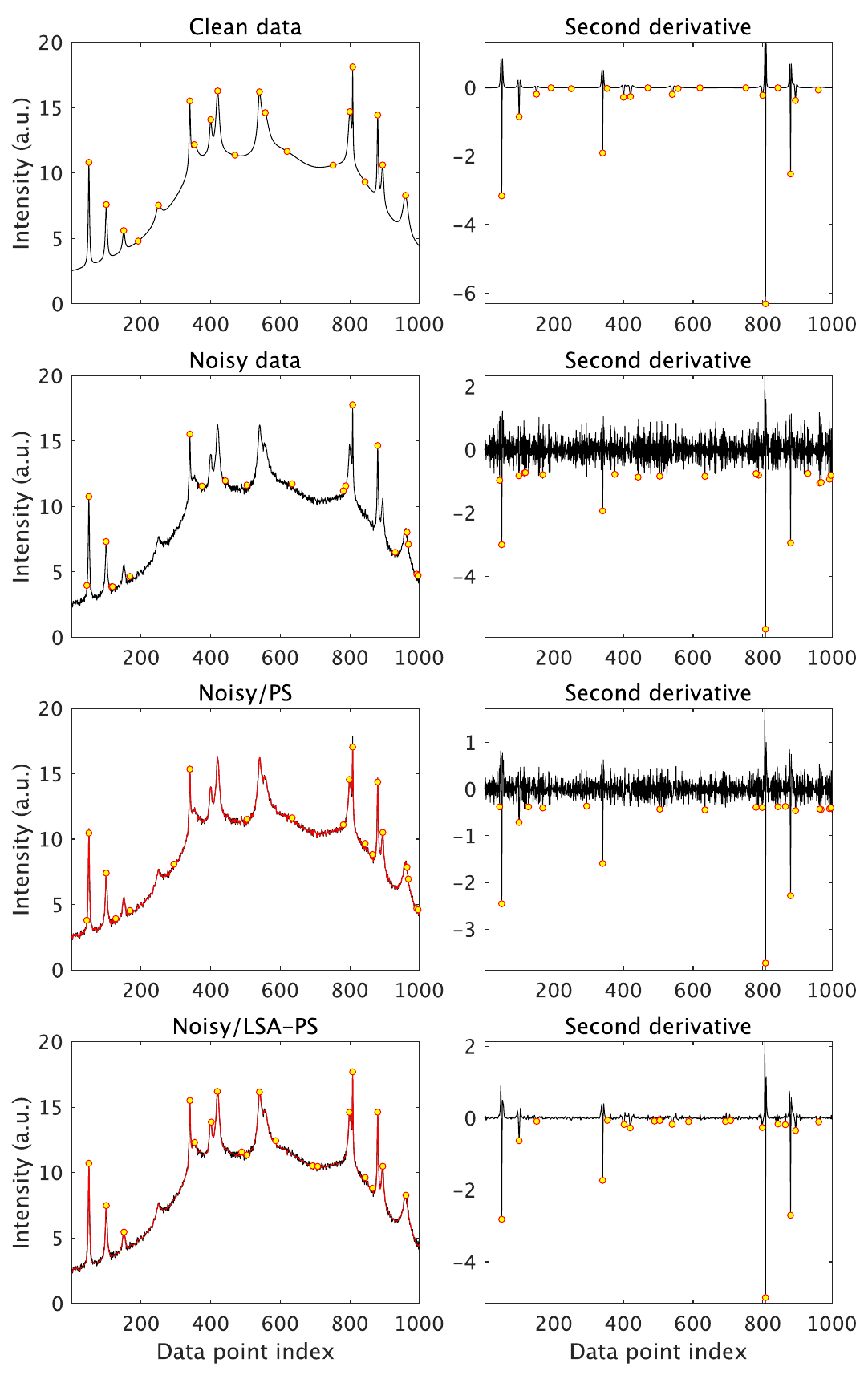}
\vspace{-0mm}
\caption{\small{From the top to the bottom: simulated data with background, noisy data, smoothed data using PS and LSA-PS with their second derivative spectra.}}
\vspace{-0mm}
\end{figure*}

\clearpage
\subsection*{Comparison of PS and LSA-PS against XRD data}
Figures 13-17 show results of PS and LSA-PS against XRD data. Figure 13-15 demonstrate how peaks are suppressed when using PS, with varying the smoothing parameter, in contrast to LSA-PS: Using LSA-PS detected prominent peaks, while PS missed some peaks, even when the smoothness was increased as some sharp peaks were suppressed. Figures 16 and 17 show results against the other two XRD data using the CVs for both PS and LSA-PS. In Figure 17, the sharp peak around 10 degree was suppressed using PS, but in contrast, using LSA-PS retained the peak intensity and also detected a more number of peaks than PS.

\begin{figure*}[h]
\centering
\includegraphics[width=18cm]{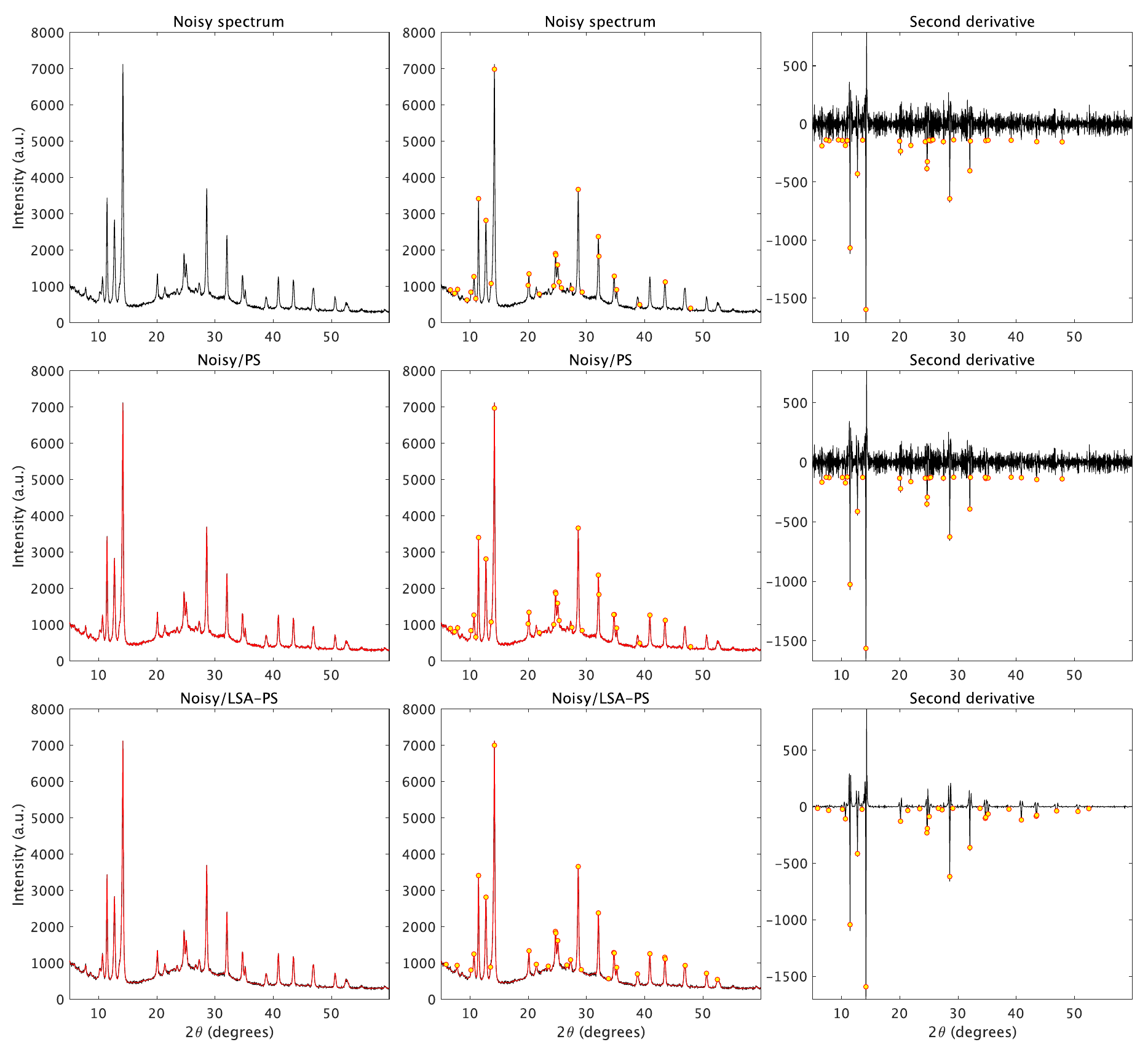}
\vspace{-2mm}
\caption{\small{From the left to the right: first sample noisy XRD data~\cite{sun2018accelerating, oviedo2019fast}, 30 detected peaks, second derivative. From the top to the bottom: without smoothing, results of PS (with estimated parameter using CV), and results of LSA-PS (with estimated parameter using Algorithm 2).}}
\vspace{-4mm}
\end{figure*}

\clearpage

\begin{figure*}
\centering
\includegraphics[width=18cm]{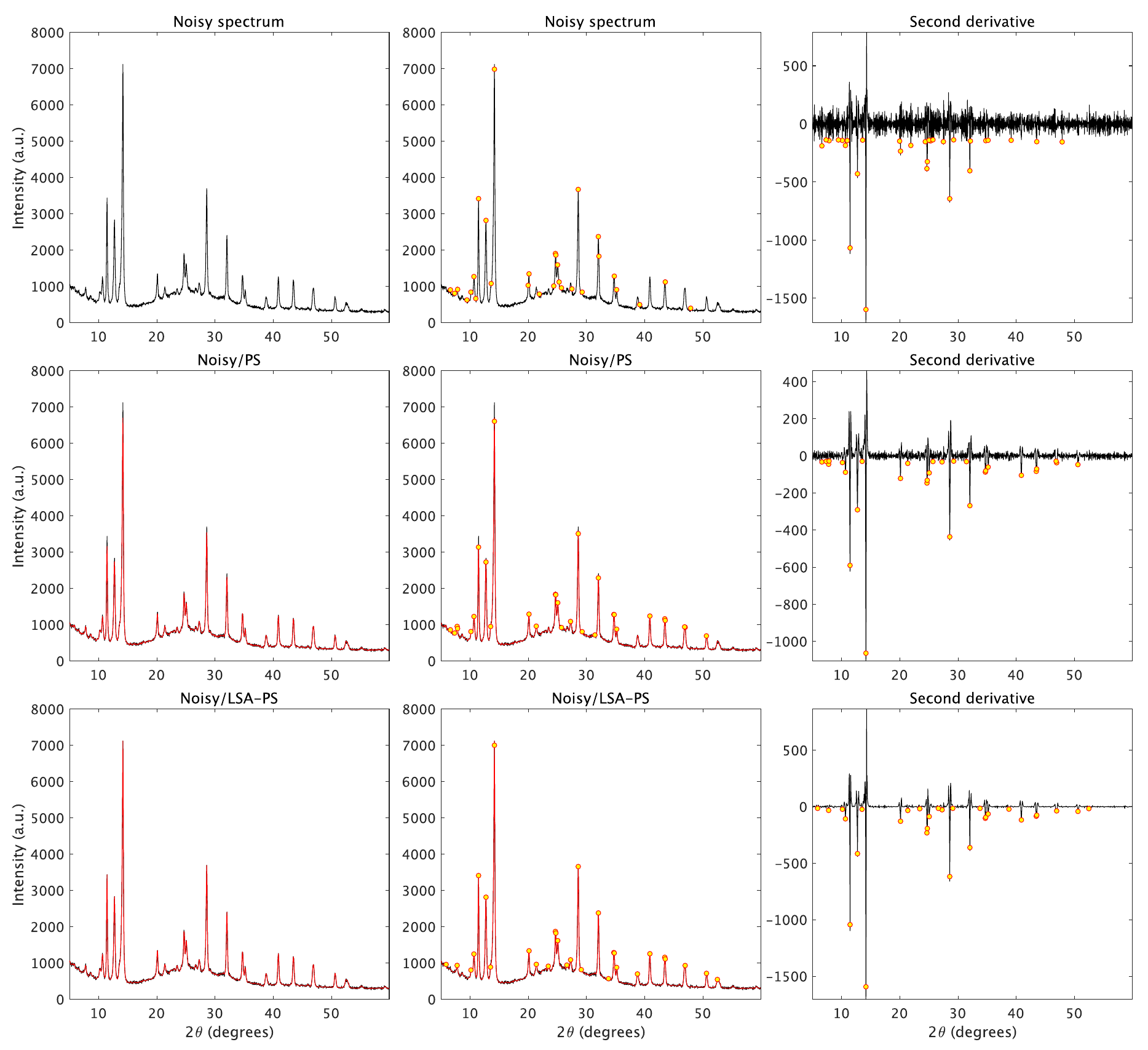}
\vspace{-2mm}
\caption{\small{From the left to the right: first sample noisy XRD data~\cite{sun2018accelerating, oviedo2019fast}, 30 detected peaks, second derivative. From the top to the bottom: without smoothing, results of PS (with {\bf{50 $\times$}} estimated parameter using the CV for PS), and results of LSA-PS (with estimated parameter using Algorithm 2).}}
\vspace{-0mm}
\end{figure*}

\clearpage

\begin{figure*}
\centering
\includegraphics[width=18cm]{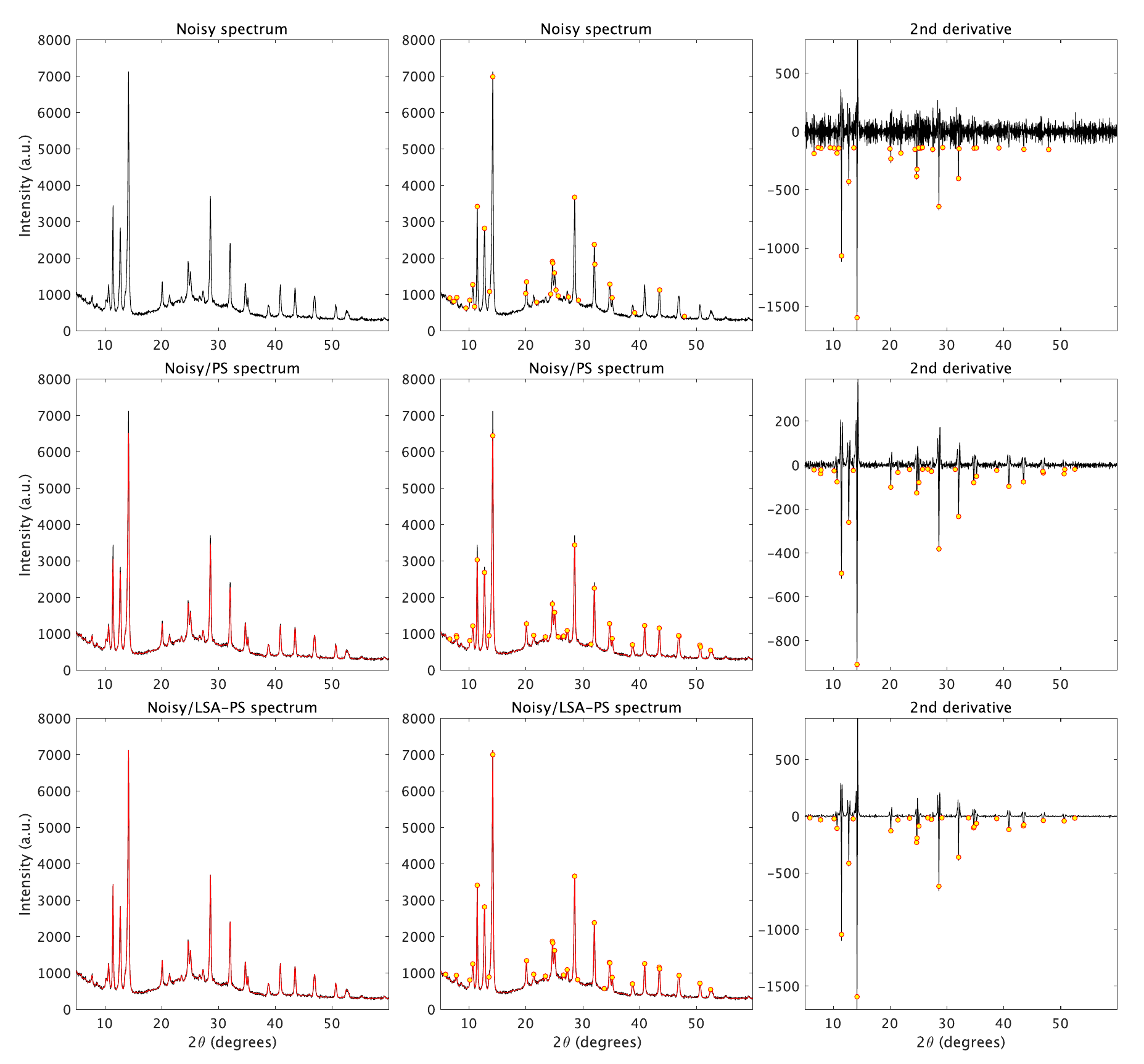}
\vspace{-2mm}
\caption{\small{From the left to the right: first sample noisy XRD data~\cite{sun2018accelerating, oviedo2019fast}, 30 detected peaks, second derivative. From the top to the bottom: without smoothing, results of PS (with {\bf{100 $\times$}} estimated parameter using the CV for PS), and results of LSA-PS (with estimated parameter using Algorithm 2).}}
\vspace{-0mm}
\end{figure*}

\clearpage

\begin{figure*}
\centering
\includegraphics[width=18cm]{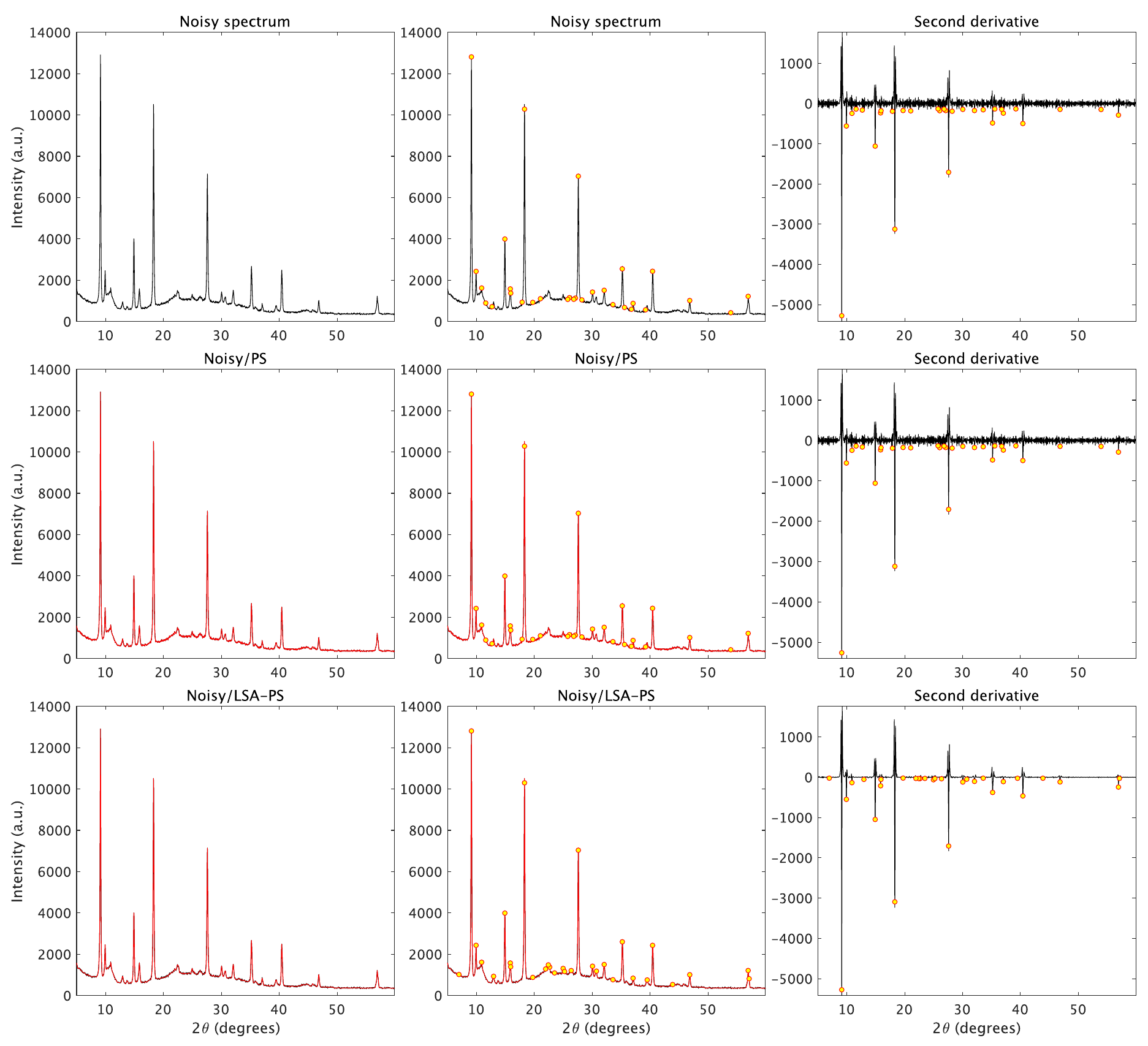}
\vspace{-2mm}
\caption{\small{From the left to the right: second sample noisy XRD data~\cite{sun2018accelerating, oviedo2019fast}, 30 detected peaks, second derivative. From the top to the bottom: without smoothing, results of PS (with estimated parameter using the CV for PS), and results of LSA-PS (with estimated parameter using Algorithm 2).}}
\vspace{-0mm}
\end{figure*}

\clearpage

\begin{figure*}
\centering
\includegraphics[width=18cm]{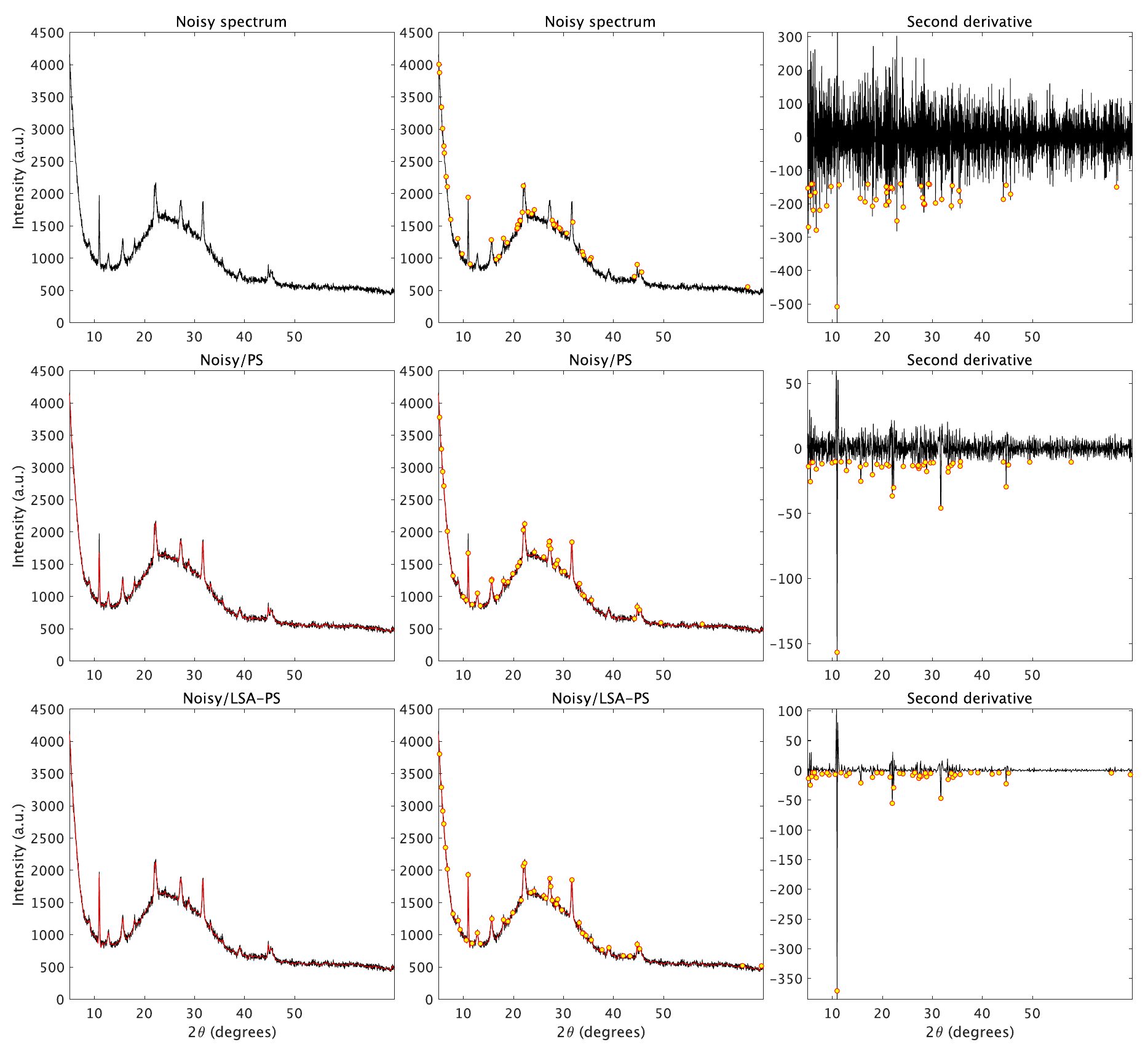}
\vspace{-2mm}
\caption{\small{From the left to the right: third sample noisy XRD data~\cite{sun2018accelerating, oviedo2019fast}, 45 detected peaks, second derivative. From the top to the bottom: without smoothing, results of PS (with estimated parameter using the CV for PS), and results of LSA-PS (with estimated parameter using Algorithm 2).}}
\vspace{-0mm}
\end{figure*}

\clearpage
\subsection*{Comparison of PS and LSA-PS against FTIR data}
Figures 18-20 show the results of PS and LSA-PS against FTIR data. Compared to PS, LSA-PS better smoothed the data while retaining sharp peaks (two peaks around 2804 ${\rm cm}^{-1}$ in Figure 18) and detected more numbers of peaks. However, after the smoothing with both PS and LSA-PS, the peaks around 3404 ${\rm cm}^{-1}$ in Figures 18-20 were not detected using second derivative. Such peaks might be detected using the other algorithms.

\begin{figure*}[h]
\centering
\includegraphics[width=18cm]{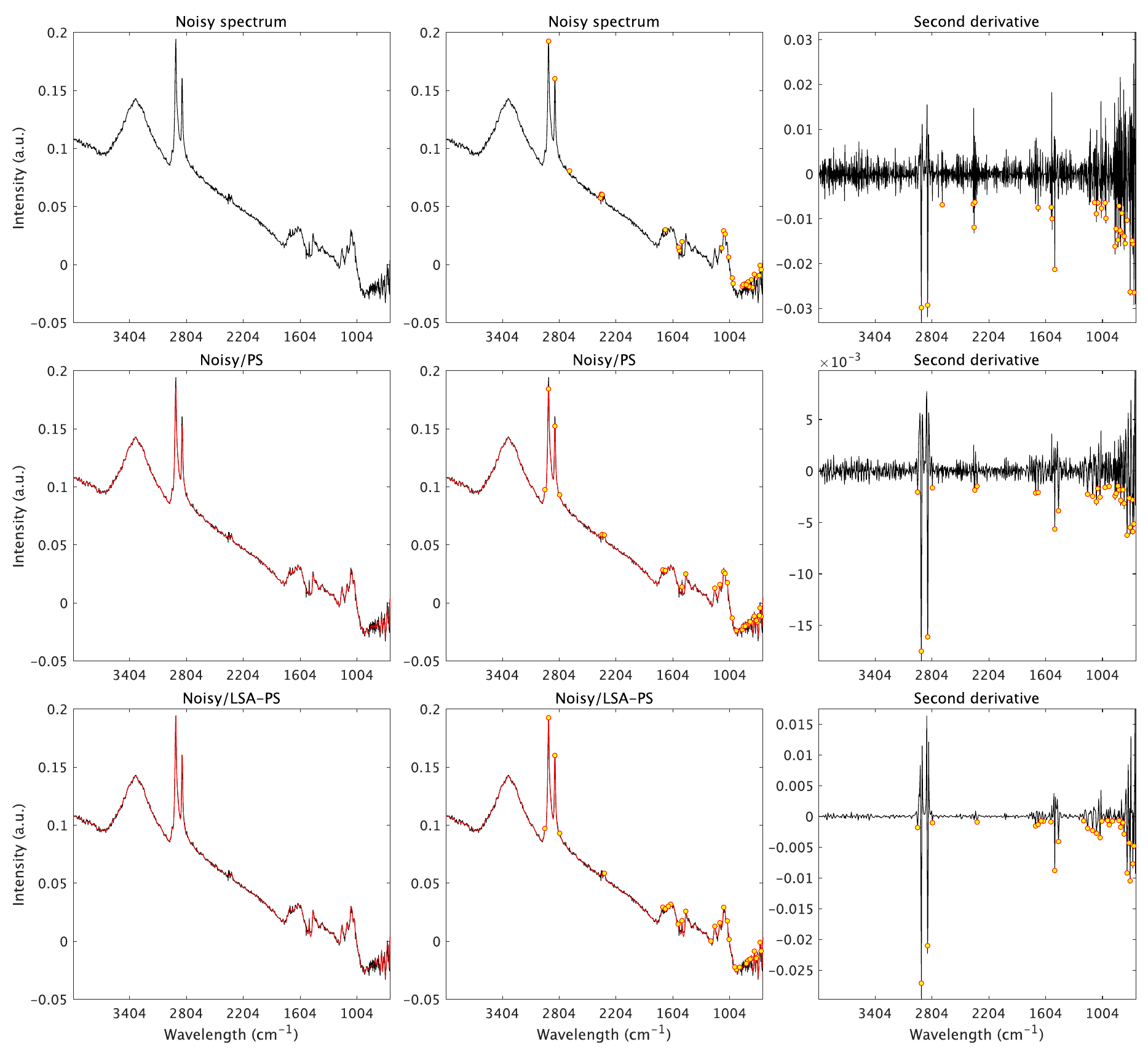}
\vspace{-2mm}
\caption{\small{From the left to the right: first sample noisy FTIR spectrum~\cite{wang2023influence, wang2023IRdataset}, 30 detected peaks, second derivative spectrum. From the top to the bottom: without smoothing, results of PS (with estimated parameter using the CV for PS), and results of LSA-PS (with estimated parameter using Algorithm 2).}}
\vspace{-0mm}
\end{figure*}

\clearpage

\begin{figure*}
\centering
\includegraphics[width=18cm]{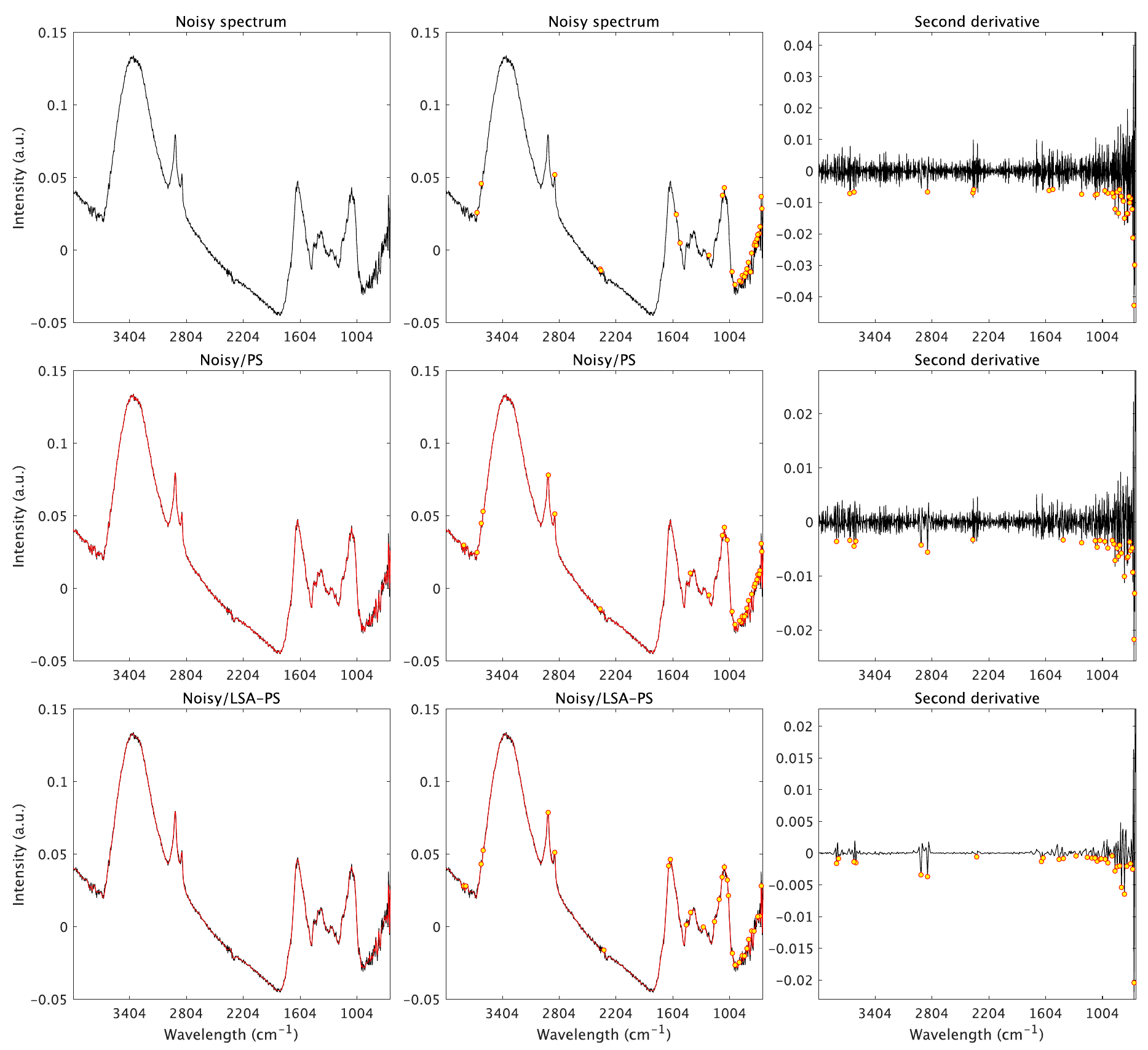}
\vspace{-2mm}
\caption{\small{From the left to the right: second sample noisy FTIR spectrum~\cite{wang2023influence, wang2023IRdataset}, 30 detected peaks, second derivative spectrum. From the top to the bottom: without smoothing, results of PS (with estimated parameter using the CV for PS), and results of LSA-PS (with estimated parameter using Algorithm 2).}}
\vspace{-0mm}
\end{figure*}

\clearpage

\begin{figure*}
\centering
\includegraphics[width=18cm]{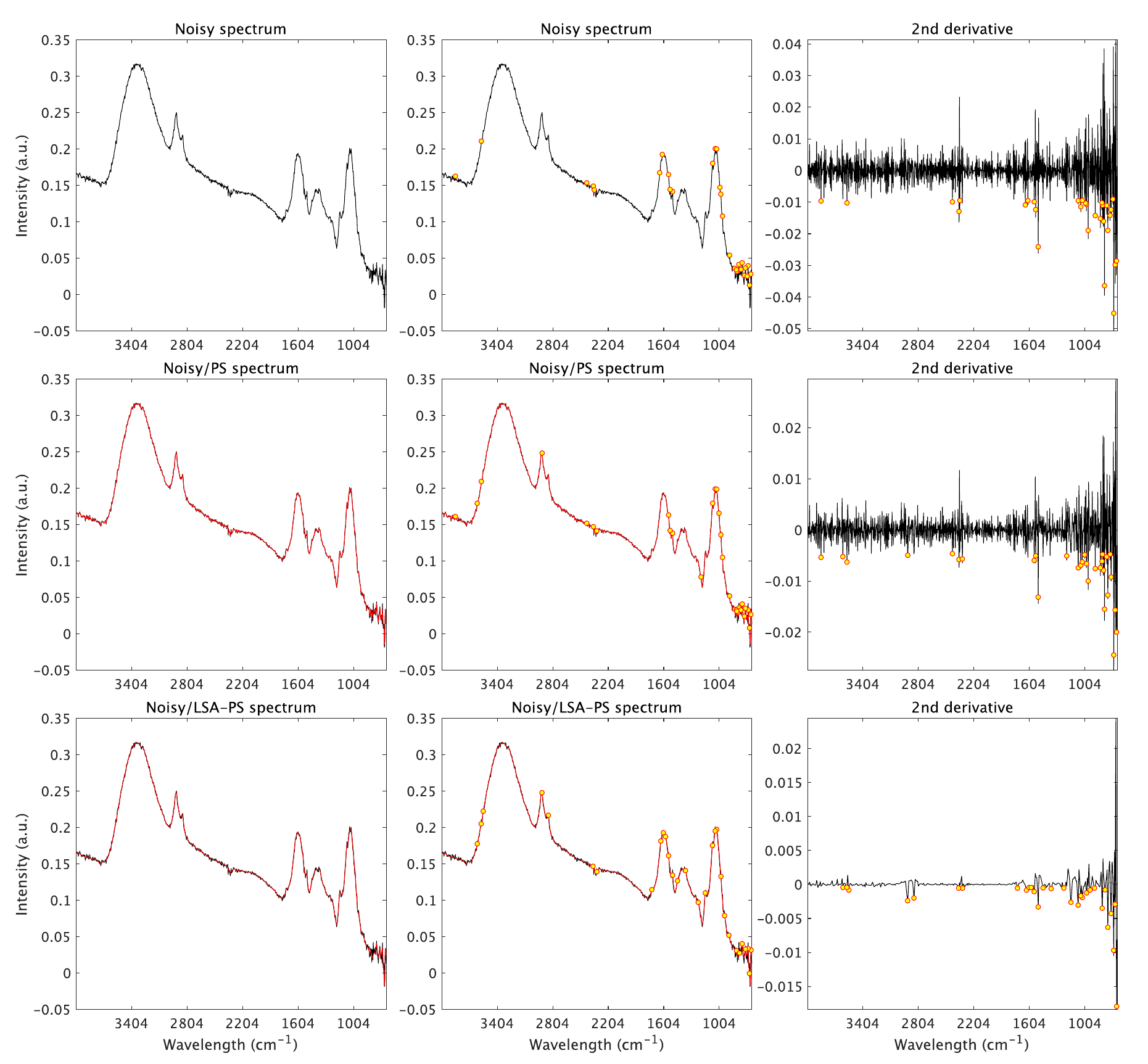}
\vspace{-2mm}
\caption{\small{From the left to the right: third sample noisy FTIR spectrum~\cite{wang2023influence, wang2023IRdataset}, 30 detected peaks, second derivative spectrum. From the top to the bottom: without smoothing, results of PS (with estimated parameter using the CV for PS), and results of LSA-PS (with estimated parameter using Algorithm 2).}}
\vspace{-0mm}
\end{figure*}

\end{document}